%% file: arxiv.tex
\documentclass[journal]{IEEEtran}
\IEEEoverridecommandlockouts
\usepackage{cite}
\usepackage{amsmath,amssymb,amsfonts}
\usepackage{algorithmic}
\usepackage{graphicx}
\usepackage{textcomp}
\usepackage[table]{xcolor}
\usepackage{booktabs}
\usepackage{tikz}
\usepackage{multirow}
\usepackage{array}
\usepackage{caption}
\usepackage{tikz}
\usepackage{hyperref}
\usepackage{orcidlink}

\usetikzlibrary{shapes.geometric, arrows, positioning}

\usetikzlibrary{arrows.meta, positioning}
\def\BibTeX{{\rm B\kern-.05em{\sc i\kern-.025em b}\kern-.08em
    T\kern-.1667em\lower.7ex\hbox{E}\kern-.125emX}}

\tikzstyle{box} = [rectangle, rounded corners, minimum width=3cm, minimum height=1cm, text centered, draw=black, fill=blue!20]
\tikzstyle{arrow} = [thick,->,>=stealth]

\begin{document}

\title{Compression in 3D Gaussian Splatting: A Survey of Methods, Trends, and Future Directions}

\author{
    Muhammad Salman Ali, 
    Chaoning Zhang \orcidlink{0000-0001-6007-6099}, 
    \IEEEmembership{Senior Member, IEEE}, 
    Marco Cagnazzo \orcidlink{0000-0001-6731-3755}, 
    \IEEEmembership{Senior Member, IEEE}, 
    Giuseppe Valenzise \orcidlink{0000-0002-5840-5743}, 
    \IEEEmembership{Senior Member, IEEE}, 
    Enzo Tartaglione \orcidlink{0000-0003-4274-8298}, 
    \IEEEmembership{Senior Member, IEEE}, and \\
    Sung-Ho Bae \orcidlink{0000-0003-2677-3186}, 
    \IEEEmembership{Member, IEEE}


\thanks{  Muhammad Salman Ali and Sung-Ho Bae are with the Department of Computer Science and Engineering, Kyung Hee University, South Korea (email: salmanali@khu.ac.kr).

        Chaoning Zhang is with School of Computer Science and Engineering, University of Electronic Science and Technology of China (UESTC), Chengdu, China. 

        Marco Cagnazzo is with LTCI, T\'el\'ecom Paris, Institut Polytechnique de Paris, France and Università degli Studi di Padova, Padua, Italy. 
        
        Giuseppe Valenzise is with the CNRS, CentraleSup\'elec, Laboratoire des Signaux et Systèmes, Université Paris-Saclay, France.
        
        Enzo Tartaglione is with LTCI, T\'el\'ecom Paris, Institut Polytechnique de Paris, France.

        }}

\maketitle
\begin{abstract}
3D Gaussian Splatting (3DGS) has recently emerged as a pioneering approach in explicit scene rendering and computer graphics. Unlike traditional neural radiance field (NeRF) methods, which typically rely on implicit, coordinate-based models to map spatial coordinates to pixel values, 3DGS utilizes millions of learnable 3D Gaussians. Its differentiable rendering technique and inherent capability for explicit scene representation and manipulation positions 3DGS as a potential game-changer for the next generation of 3D reconstruction and representation technologies. This enables 3DGS to deliver real-time rendering speeds while offering unparalleled editability levels. However, despite its advantages, 3DGS suffers from substantial memory and storage requirements, posing challenges for deployment on resource-constrained devices. In this survey, we provide a comprehensive overview focusing on the scalability and compression of 3DGS. We begin with a detailed background overview of 3DGS, followed by a structured taxonomy of existing compression methods. Additionally, we analyze and compare current methods from the topological perspective, evaluating their strengths and limitations in terms of fidelity, compression ratios, and computational efficiency. Furthermore, we explore how advancements in efficient NeRF representations can inspire future developments in 3DGS optimization. Finally, we conclude with current research challenges and highlight key directions for future exploration.
\end{abstract}

\section{Introduction}
\IEEEPARstart{T}{ransforming} a collection of views, images, or video capturing a scene into a 3D model such that computers can process it is the goal of image-based 3D scene reconstruction. This complex and enduring problem is crucial for enabling machines to understand the complexities of real-world environments, paving the way to a diverse range of applications including 3D animation and modeling,  navigation of robots, scene preservation, virtual/augmented reality, and autonomous driving~\cite{kalkofen2008comprehensible,patney2016towards,albert2017latency}. The evolution of 3D scene reconstruction precedes the rise of deep learning, with initial efforts focusing on light fields and fundamental scene reconstruction techniques~\cite{gortler2023lumigraph,levoy2023light,buehler2023unstructured}. However, these early works faced limitations due to their dependence on dense sampling and structured capture, which presented significant challenges in managing complex scenes and lighting variations. The introduction of structure-from-motion~\cite{snavely2006photo,goesele2007multi} and the subsequent enhancements in multiview stereo algorithms offered a more resilient foundation for 3D scene reconstruction. However, these approaches encountered difficulties in synthesizing novel views and lacked alignment with deep scene understanding models.

Before Gaussian Splatting gained its popularity for the task of 3D reconstruction, NeRFs~\cite{mildenhall2020nerf} constituted as a go-to method, representing a significant breakthrough in this advancement. Using fully connected neural networks, NeRFs facilitate the direct mapping of spatial coordinates to color and density. The success of NeRFs lies in their capability to generate continuous, volumetric scene functions, yielding results with remarkable detail and realism~\cite{liao2024ov,lin2025dynamic,sheng2024open,zhu2024dfie3d,ding2024ray}. However, like any emerging technology, this implementation incurs certain costs.
\begin{enumerate}
    \item Computational Complexity: NeRF-based methods have significantly high computational complexity~\cite{chen2022tensorf,garbin2021fastnerf,takikawa2021neural}, often requiring long training times and significant resources for rendering, particularly for high-resolution outputs. 
    \item Scene Editability: Performing implicit scenes can be challenging, as modifying the neural network’s weights does not directly correspond to changes in the geometry or appearance of the scene~\cite{ali2024elmgs,qian20233dgs,lee2023compact}.
\end{enumerate}

\input{Tables/Taxonomy_Table}
\input{Display_Figures/taxonomy}



Although NeRFs are adept at producing photorealistic images, there is a growing need for faster and more efficient rendering techniques, especially for applications where low latency is essential. 3D Gaussian Splatting (3DGS) solves this problem by employing millions of learnable 3D Gaussians in space for explicit scene representation in scene modeling. 3DGS employs an explicit representation and a highly parallelized rasterization approach, which facilitates more efficient computation and rendering, unlike implicit coordinate-based models~\cite{henzler2019escaping,sitzmann2019deepvoxels,mildenhall2020nerf}. The innovation of 3DGS lies in its integration of differentiable pipelines and point-based rendering techniques~\cite{pfister2000surfels,zwicker2001surface,ren2002object,botsch2005high}. Modeling the scenes with learnable 3D Gaussians preserves the robust overfitting capabilities of continuous volumetric radiance fields required for high-quality image synthesis. Simultaneously, it circumvents the computational complexity of NeRF-based methods, such as the computationally expensive ray marching process and redundant calculations in unoccupied space~\cite{10757420}.

The introduction of 3DGS signifies more than just a technical leap forward; it represents a fundamental change in the approach to scene representation and rendering within computer vision and graphics. By facilitating real-time rendering capabilities while maintaining high visual fidelity, 3DGS paves the way for several applications ranging from virtual reality and augmented reality to real-time cinematic rendering and beyond~\cite{jiang2024vr}. This advancement promises to not only enhance current applications but also unlock new ones that were previously hindered by computational limitations~\cite{liu2024georgs,10879794}. Furthermore, the explicit scene representation provided by 3DGS offers unparalleled flexibility in managing objects and scene dynamics, which is essential to handle complex scenarios with intricate geometries and diverse lighting conditions~\cite{chabra2020deep, wang2021learning}. 
This high degree of editability combined with the efficiency of both the training and rendering processes enables 3DGS to have a deep impact on future advancements in different domains~\cite{10900457}. As a relatively recent emergence, within less than a year, the multitude of works on 3DGS underscores its broad applicability across diverse domains such as robotics~\cite{yan2024gs,keetha2024splatam,matsuki2024gaussian,yugay2023gaussian,huang2024photo}, avatars~\cite{li2024animatable,hu2024gauhuman,lei2024gart,yuan2024gavatar,hu2025tgavatar}, endoscopic scene reconstruction~\cite{huang2024endo,liu2024endogaussian,zhao2024hfgs,wang2024endogslam}, and physics~\cite{xie2024physgaussian,liu2024physics3d,borycki2024gasp,huang2024dreamphysics,zhang2024physdreamer}. 

Compared with NeRFs, 3DGS has the advantage of faster rendering speed but at the cost of higher demand of memory with the need to store millions of Gaussians. This limits their application in resource-constrained devices, like VR/AR or game environments. Therefore, there has been a notable increase in research activities focused on compressing 3DGS scenes. With the development of numerous novel compression methods, these efforts have resulted in significant advancements in compression ratios. Consequently, there is a pressing need for a timely review and summary of these representative methods for 3DGS compression. Such a review would help researchers grasp the overall landscape of 3DGS compression, providing a comprehensive and structured overview of current achievements and major challenges in the field.

\input{Display_Figures/3DGS_working}

\noindent
\textbf{Scope:} This survey will explore the compression-specific design of the 3DGS architecture, covering various modules including but not limited to densification of Gaussians, pruning, vector quantization, scalar quantization, Gaussian structure, and point cloud compression.

\noindent
\textbf{Related Surveys: }A few 3DGS surveys exist~\cite{DBLP:journals/corr/abs-2401-03890,fei20243d, bao20243d,wu2024recent}. However, to the best of our knowledge, only one survey focuses on the compression of 3DGS. 3DGS.zip~\cite{bagdasarian20243dgs} provides an overview of existing compression techniques, with its primary contribution being the establishment of a unified evaluation standard.  In contrast, our work systematically categorizes different compression methods within a proposed taxonomy (Table~\ref{tab:Taxonomy}, Figure~\ref{fig:taxonomy}) of structured and unstructured approaches, analyzing their distinctions and associated challenges. Additionally, we offer insights into future research directions for both structured and unstructured techniques, as well as perspectives from point cloud and NeRF compression, which are missing in the prior survey.  Furthermore, our work is the first comprehensive study to provide an in-depth discussion on the efficiency (rendering speed) of different compression methods.  

\noindent
\textbf{Highlighted Features: } The major features of this survey include (1) Highlighting Compression Potential: This survey underscores the potential of 3DGS to be highly compressed with minimal quality loss. We demonstrate that 3DGS is compatible with various compression methodologies and their combinations. For the first time, we provide a comprehensive understanding of different 3DGS compression methodologies from a topological perspective. (2) Key Concepts and Improvements: We discuss the essential concepts involved in compressing 3DGS and outline strategies for further improvements. This includes an in-depth examination of current techniques and their impact on compression efficiency and quality preservation. (3) Guidelines for Future Research: Drawing from recent advancements, we extract pivotal ideas from state-of-the-art compression methodologies. Based on these insights, we propose a set of guidelines for future research to enhance 3DGS compression pipelines. This includes best practices and innovative approaches to be incorporated in future studies.

As the pioneering attempt to present a comprehensive survey on 3DGS compression, our survey aims to help the readers of interest quickly grasp its development. Overall, the contributions of our survey are summarized as follows:

(1) The paper offers a detailed taxonomy of 3DGS compression methods, categorizing them into structured and unstructured approaches, and provides an in-depth analysis of their methodologies, performance trade-offs, and limitations. 

(2) We systematically analyze existing compression techniques, highlighting key challenges such as scalability constraints in large-scale scenes, dependence on vector quantization, suboptimal loss function designs, and limitations in deploying 3DGS on resource-constrained hardware.

(3) We provide insights for advancements including scalar quantization for efficient hardware compatibility, hybrid frameworks that integrate structured and unstructured compression strategies, and leveraging insights from NeRF and point cloud compression to enhance performance and adaptability.



The remainder of the paper is structured as follows: Section II presents a detailed background on 3DGS and its working principles. Section III introduces the problem statement for 3DGS compression. Sections IV and V discuss unstructured and structured compression methods, respectively. Section VI provides a comparative analysis of both techniques. Section VII explores future directions inspired by NeRFs and point cloud compression. Finally, Section VIII concludes the paper.  
\input{Sections/Preliminaries_arxiv}

 \section{3DGS Compression}
 \label{sec:compression}

3DGS faces significant scalability challenges compared to NeRFs. While NeRFs require only the storage of weight parameters for a multilayer perceptron (MLP), 3DGS necessitates storing the parameters of millions of Gaussians per scene. This issue becomes especially critical in large, complex scenes, where computational and memory demands increase significantly. The number of Gaussians is directly proportional to storage and computational complexity and inversely proportional to rendering efficiency. Therefore, optimizing memory usage and computational efficiency for both storage and rendering is essential to improve the scalability of 3DGS-based methods.

\noindent
\textbf{3DGS Attributes: }The input signal to be compressed consists of $N$ Gaussians, each characterized by multiple attributes: $3\times1$ position vectors ($\mu$), $3\times1$  scale vector and $4\times1$  rotation quaternion vector, scalar opacity values and spherical harmonics (SH) coefficients for view-dependent RGB color modeling. For degree $3$ SH, this requires $48$ coefficients per Gaussian ($16$ per RGB channel). Each parameter has a bit depth of 32-bit floating point.

\noindent
\textbf{Evaluation Metrics:} The compression cost is quantified either as the bit count of the compressed representation or its compression ratio compared to the baseline 3DGS-30k (trained for $30,000$ iterations). The performance evaluation of 3DGS compression relies on image-based fidelity metrics such as PSNR (Peak Signal-to-Noise Ratio), SSIM (Structural Similarity Index Measure), and LPIPS (Learned Perceptual Image Patch Similarity), and its computational efficiency is measured in terms of Frames per second (FPS). 

\noindent
\textbf{Datasets:} Similar to NeRF-based rendering methods, 3DGS-based methods are commonly evaluated on 9 scenes from Mip-NeRF360~\cite{barron2022mip}, which includes both indoor and outdoor scenes, two scenes from Tanks\&Temples~\cite{knapitsch2017tanks}, and the Deep Blending~\cite{hedman2018deep} dataset. Figure~\ref{fig:dataset} shows a sample image of each scene from all the datasets. To ensure consistent benchmarking studies typically adhere to the train-test split used in Mip-NeRF360~\cite{barron2022mip} and 3DGS where every 8th image must be selected for testing.

\input{Tables/Unstruct_table}

\noindent
 \textbf{Categorization of 3DGS Compression Methods:}
There has been a plethora of works focusing on the compression of 3DGS. In contrast to the structured feature grids used in NeRF-based methods, the 3D Gaussians employed in 3DGS are sparse and lack organization, which causes significant difficulties in establishing structural relations~\cite{chen2024hac}. Consequently, compression strategies for 3DGS can be categorized into two groups: \newline i) \textbf{Unstructured Compression:} Those primarily concerned with compressing the "value" of model parameters $N$, utilizing methods like pruning~\cite{fan2023lightgaussian,lee2023compact}, quantization~\cite{fan2023lightgaussian,lee2023compact,navaneet2023compact3d,niedermayr2024compressed}, and entropy constraints~\cite{girish2023eagles} without considering the relationship between the Gaussians. \newline ii) \textbf{Structured Compression:} those exploring compression techniques that consider the relationships between Gaussians~\cite{lu2023scaffold,chen2024hac}.

Figure~\ref{fig:taxonomy} presents a detailed taxonomy of 3DGS compression methods, while Table~\ref{tab:Taxonomy} lists representative publications categorized according to their taxonomy classification. The following sections provide an overview of unstructured and structured compression methods, discussing their variations and the challenges associated with each approach.

\vspace{-0.15cm}
\section{Unstructured 3DGS Compression} 
Compression methods built on top of the baseline 3DGS that exploit the sparse nature of Gaussians without altering the fundamental structure of 3D Gaussians or considering their interrelationships fall into this category. These approaches apply techniques such as pruning, quantization, and entropy coding within the existing 3DGS framework, making minimal modifications to the underlying architecture.  The primary objective of these methods is to reduce memory and computational costs while preserving the core advantages of 3DGS, ensuring efficient storage and faster processing without degrading scene representation quality. In this section, based on the taxonomy presented in Figure~\ref{fig:taxonomy} and Table~\ref{tab:Taxonomy}, we provide an in-depth analysis of unstructured compression methods, discussing their performance, challenges, and potential future directions.  

\subsection{Pruning}
Pruning techniques in 3DGS aim to reduce the number of Gaussians resulting in storage optimization and rendering efficiency while maintaining rendering fidelity. These approaches target different Gaussian attributes, utilizing structural, statistical, and learned information to optimize scene representation. The pruning strategies can be broadly classified into size-based, gradient-based, opacity-based, spatial-based, and significance-scoring-based techniques.

\noindent
\textbf{Size-based Pruning:} This method eliminates Gaussians that are structurally redundant due to their small size, as they contribute minimally to scene reconstruction. CompGS~\cite{navaneet2023compact3d} and Papantonakis et al.~\cite{papantonakis2024reducing} apply size-based pruning to remove such Gaussians, improving efficiency while maintaining reconstruction quality.

\noindent
\textbf{Gradient-based Pruning:} This method removes Gaussians that contribute minimally to optimization by evaluating the magnitude of gradients associated with each Gaussian during training. EfficientGS~\cite{liu2024efficientgs} utilizes cumulative gradient sum analysis to halt unnecessary densification and applies pruning to eliminate redundant Gaussians. GDGS~\cite{gong2024gdgs} reduces Gaussian density by modeling scene gradients, enhancing compactness. Trimming the Fat~\cite{salman2024trimming} and ELMGS~\cite{ali2024elmgs} applies gradient-based pruning, removing 75\% of Gaussians while maintaining high visual quality. Kim et al.~\cite{kim2024color} extend this approach by incorporating SH gradients alongside positional gradients, further refining the pruning process.

\input{Display_Figures/unstruct_fps_vs_psnr}

\noindent
\textbf{Opacity-based Pruning:} This method eliminates Gaussians with low opacity values, as they contribute minimally to scene reconstruction. Trimming the Fat~\cite{salman2024trimming} and ELMGS~\cite{ali2024elmgs} apply opacity-based pruning alongside gradient-based pruning, effectively removing floaters and redundant Gaussians. CompGS~\cite{navaneet2023compact3d}, in combination with size-based pruning, employs a learnable opacity masking approach, dynamically removing Gaussians with persistently low opacity throughout training. Figure~\ref{fig:opacity_gradient_pruning_comparison} presents a qualitative comparison between pruning based solely on opacity and pruning that incorporates both opacity and gradient information. The results highlight the effectiveness of opacity and gradient-informed pruning in preserving scene details while achieving better compression efficiency.

\noindent
\textbf{Significance-based pruning:} This approach incorporates explicit scoring functions to regulate pruning decisions, preventing excessive removal of Gaussians that could degrade scene fidelity. LightGaussian~\cite{fan2023lightgaussian} employs a significance-driven pruning approach that evaluates each Gaussian’s contribution to rendering based on its projection onto camera viewpoints, ensuring minimal perceptual degradation. The significance score is computed from the frequency of Gaussian intersections with rays across all training views. EAGLES~\cite{girish2023eagles} adopts a coarse-to-fine pruning strategy, eliminating Gaussians with the least contribution to reconstruction quality, thereby enhancing training and inference speeds. Papantonakis et al.~\cite{papantonakis2024reducing} combine size-based and significance-based pruning, dynamically adapting SH coefficients to remove structurally redundant Gaussians. SafeguardGS~\cite{lee2024safeguardgs} introduces a pruning score function to ensure optimal Gaussian selection, mitigating the risk of catastrophic scene degradation. LP-3DGS~\cite{zhang2024lp} adopts a trainable binary mask approach that automatically determines the optimal pruning ratio, leveraging Gumbel-Sigmoid-based gradient approximation to maintain compatibility with existing 3DGS training pipelines.

\input{Display_Figures/unstruct_mem_vs_psnr}

\noindent
\textbf{Spatial-based Pruning:} This method removes Gaussians based on scene location, preserving detail in important regions while reducing redundancy elsewhere. PUP 3D-GS~\cite{hanson2024pup} employs a second-order reconstruction error approximation to selectively prune Gaussians with minimal impact on scene reconstruction. RTGS~\cite{lin2024rtgs}, using a foveated rendering (FR)~\cite{guenter2012foveated,patney2016towards} approach for Point-Based Neural Rendering (PBNR)~\cite{kerbl20233d}, prunes Gaussians based on pixel eccentricity~\cite{wandell1995foundations}, maintaining high-density in critical regions while sparsifying peripheral areas. This approach optimizes memory usage and rendering speed while preserving perceptual consistency, making it particularly effective for real-time rendering, VR, and AR applications.


By integrating these diverse pruning methodologies, 3DGS achieves substantial memory reduction and enhanced rendering speeds, making possible real-time Gaussian rendering across various applications, including VR/AR~\cite{zhai2024splatloc}, mobile deployment~\cite{lin2024rtgs}, and autonomous systems~\cite{zhou2024drivinggaussian}. 
However, even after removing redundant Gaussians through pruning, millions of Gaussians are still required for accurate scene reconstruction. To further reduce the memory and storage complexity of these essential Gaussians, quantization techniques are employed.

\subsection{Quantization}
Quantization plays a crucial role in 3DGS compression by reducing the bit precision of Gaussian attributes while maintaining rendering fidelity. By encoding Gaussian parameters more compactly, quantization significantly decreases storage requirements and computational costs. Several works integrate quantization within their compression pipelines, either in isolation or combined with pruning, to achieve higher efficiency. Quantization can be broadly categorized into scalar quantization and vector quantization. Scalar quantization compresses individual attributes independently, while vector quantization groups multiple attributes into a shared codebook, enabling more efficient compression and reduced storage overhead.

\noindent
\textbf{Vector Quantization (VQ):} VQ is widely used in 3DGS compression due to its higher compression efficiency, achieved by clustering similar Gaussians and encoding them through compact indices. LightGaussian~\cite{fan2023lightgaussian} and CompGS~\cite{navaneet2023compact3d} employ codebook-based vector quantization to identify shared Gaussian parameters, further compressing indices via run-length encoding. Niedermayr et al.~\cite{niedermayr2024compressed} introduce sensitivity-aware vector clustering combined with quantization-aware training, optimizing directional color and Gaussian parameters. EAGLES~\cite{girish2023eagles} integrates quantized embeddings, significantly reducing per-point memory requirements while accelerating training and inference.

\noindent
\textbf{Scalar Quantization (SQ):} Although VQ is widely used in 3DGS-based methods, recent advances suggest that SQ offers better hardware efficiency, particularly for deployment in edge devices~\cite{gholami2022survey}. ELMGS~\cite{ali2024elmgs} used learned step-size-based uniform quantization~\cite{esser2019learned} for quantization. However, uniform SQ suffers from a performance drop at lower bit-depths due to the non-uniform distribution of 3DGS attributes such as opacity as seen in Figure~\ref{fig:opacity_distribution}~\cite{esser2019learned}. The distribution of opacity in 3DGS is highly non-uniform, with peaks at both lower and higher opacity levels. This underscores the necessity for non-uniform SQ methods specifically tailored for 3DGS quantization, ensuring better preservation of scene fidelity even at extremely low bit-depths.

\subsection{Entropy Encoding}
After quantization, entropy encoding further compresses Gaussian attributes by eliminating redundancy at the storage level, ensuring a compact representation. CompGS, ELMGS, and Niedermayr et al. apply run-length encoding and entropy-based compression to minimize storage overhead. RDO-Gaussian~\cite{wang2024end} is among the first to introduce an end-to-end rate-distortion framework, dynamically adjusting compression based on a quality-loss trade-off. Morgenstern et al.~\cite{morgenstern2023compact} propose a 2D-grid-based entropy encoding that efficiently organizes Gaussian attributes, enabling parallelized encoding and decoding during rendering.

Entropy encoding serves as a crucial final step in 3DGS compression, significantly enhancing compression efficiency by further reducing storage redundancy. However, it introduces additional computational complexity during decoding. Methods such as RDO-Gaussian, which integrate entropy encoding with quantization and pruning, offer the most efficient end-to-end compression pipelines, balancing compression ratio, and computational complexity.

\input{Sections/Unstructured_Compression}

\noindent
\textbf{Challenges and Future Direction}
Despite substantial progress in compressing Gaussian splats, the majority of research has primarily concentrated on reducing the storage footprint of 3DGS. A critical challenge that remains unaddressed is the densification of 3DGS, particularly for large scenes like those in the Mip-NeRF dataset~\cite{barron2022mip}, such as the \texttt{garden} and \texttt{bicycle} scenes, which can demand up to 15GB of GPU memory for training and rendering. This level of scaling is unsustainable for large-scale scenes, indicating a pressing need for unstructured 3DGS compression techniques to tackle this issue.

Furthermore, most of the existing compression methods rely on vector quantization. However, recent advances in model compression for deep learning have demonstrated that scalar quantization is more favorable and easier to implement in hardware, especially for low-powered edge devices~\cite{gholami2022survey,nascimento2023hyperblock}. Therefore, developing specialized scalar quantization techniques for efficiently rendering Gaussian splats on such devices should be a priority.

Additionally, none of the current compression works have investigated the behavior of loss functions in relation to different compression methods. In image compression, it has been shown that altering the loss function can significantly impact compression performance~\cite{ali2024towards}. Thus, future research should explore the effects of various loss functions, such as perceptual loss and edge-based loss, on the performance of 3DGS compression. This could lead to more efficient and perceptually optimized compression techniques for 3DGS.

\input{Tables/struct_table}

\input{Sections/Structured_Compression}

\noindent
\textbf{Challenges and Future Direction: }
Structured compression organizes sparse Gaussians, offering significant memory footprint reduction as discussed in earlier sections. However, the added complexity from hashing~\cite{chen2024hac}, offsetting~\cite{lu2023scaffold}, encoding, and decoding~\cite{liu2024compgs} introduces computational overhead, which impacts rendering efficiency. Unlike unstructured compression, where compression gains typically lead to faster rendering, structured methods do not inherently translate compression efficiency into rendering efficiency. 


One challenge with structured compression lies in the diversity of methodologies employed by different approaches, which complicates their scalability and standardization. Unlike unstructured compression methods, structured techniques cannot be easily integrated into the baseline 3DGS as plug-and-play solutions.

To advance the field, future research should aim to merge insights from unstructured compression into structured approaches, particularly to enhance rendering speed. Given that techniques inspired by LIC have already influenced 3DGS, it is likely that future work will delve deeper into exploring relationships among Gaussians, developing strategies for effective grouping and combining of Gaussians, and optimizing their encoding and decoding processes for greater efficiency and effectiveness.

\begin{figure*}[t]
    \centering
    \includegraphics[width=\linewidth]{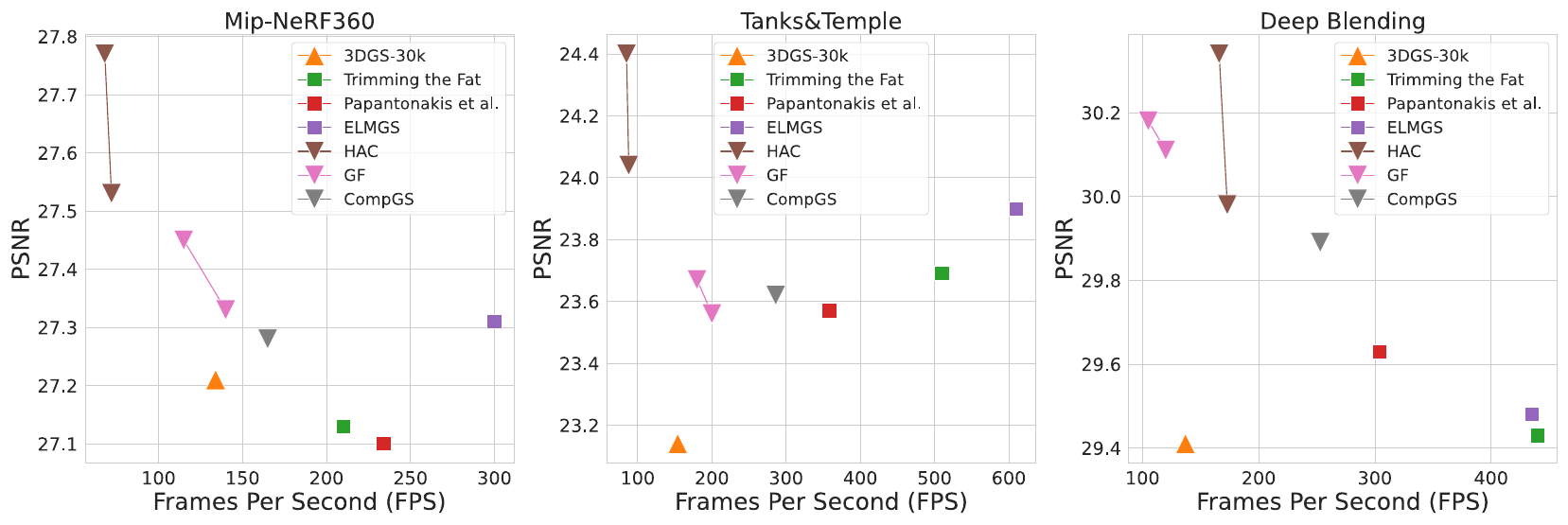}
    \caption{Comparison of PSNR and FPS across structured and unstructured compression techniques benchmarked against the baseline 3DGS. The box represents unstructured compression methods, while the upward triangle denotes the baseline 3DGS-30k. The downward triangle indicates structured compression methods. FPS values are measured on a single NVIDIA A40 GPU.}
    \label{fig:fpsvspsnr}
\end{figure*}

\section{Structured Compression vs Unstructured Compression}
\label{sec:struc_vs_unstruc}

Both structured and unstructured compression techniques significantly reduce the memory footprint of 3DGS, with the choice of method depending on application-specific needs. 

\noindent
\textbf{Fidelity:} Structured compression methods enhance PSNR and SSIM by imposing organization within the sparse Gaussian framework, as seen in ScaffoldGS and HAC-highrate, which achieve PSNR values of 30.21 on Deep Blending and 27.77 on Mip-NeRF360. Unstructured methods, in contrast, preserve the flexibility of baseline 3DGS-30K, offering a more direct adaptation while maintaining competitive fidelity.

\noindent
\textbf{Compression Ratio: } Structured techniques outperform unstructured methods by achieving higher compression rates while retaining visual quality. CompGS (lowrate) and ContextGS (lowrate) achieve 70$\times$ and 113$\times$  compression on Mip-NeRF360 and Deep Blending, respectively, making them highly efficient for memory-constrained settings. Unstructured methods, while not matching these extreme ratios, still achieve significant memory reductions without additional constraints, offering a better balance between compression and flexibility. Figure~\ref{fig:memvspsnr} shows the compression-performance tradeoff of structured and unstructured compression methods.

\noindent
\textbf{FPS: }Structured methods remain comparable to the 3DGS-30K baseline, whereas unstructured techniques achieve 4$\times$  to 5$\times$  FPS improvements, as shown in Figure~\ref{fig:fpsvspsnr}. This makes unstructured compression preferable for real-time and low-power applications, where efficiency is critical. While structured methods introduce additional parameters that can slightly affect rendering speed, they still maintain a high FPS suitable for interactive applications.

\noindent
\textbf{Challenges and Future Direction: }Despite these advancements, challenges remain. Vector quantization, while effective for compression, is computationally expensive compared to scalar quantization~\cite{ali2024elmgs}. Additionally, optimizing structured methods for low-power edge devices remains underexplored. Adaptive rasterization strategies, such as those proposed by Niedermayr et al.~\cite{niedermayr2024compressed}, could enhance rendering efficiency while refining loss functions could further improve 3DGS quality~\cite{ali2024towards}. Addressing these issues will enable more efficient and high-quality 3DGS compression techniques, benefiting both structured and unstructured approaches.

\section{Future Works and Direction}
\label{sec:future_works}
\subsection{Inspiration from NeRFs}
In the years between 2020 and 2023, we have witnessed rapid advancements in NeRFs, as moving through a very steep trajectory, multiple compression techniques have significantly impacted the design of newer generation NeRFs. Recent work on NeRF compression, such as~\cite{tancik2022block} demonstrated how neural network-based approaches can efficiently represent large-scale 3D scenes with compact latent representations. This paradigm suggests that 3DGS compression could similarly benefit from end-to-end deep learning techniques that directly optimize 3D Gaussian representations, rather than relying merely on traditional data compression methods. By learning more efficient latent space representations of 3D objects, neural network-based models like NeRFs have shown the potential to dramatically reduce the storage and transmission requirements for high-fidelity 3D scenes, which could inspire similar methods for compressing 3D Gaussian components.

One of the key contributions of NeRFs to the field of compression is the use of hierarchical representations~\cite{tang2022compressible,zheng2024hpc}. In NeRFs, the scene is encoded in multiple levels of resolution, progressively refining the details of the 3D structure and lighting~\cite{di2025boost}. This hierarchical approach can be applied to 3D Gaussian Splitting by allowing for multi-level Gaussian components to be represented at varying degrees of detail, depending on the importance of different regions of the 3D space. The concept of progressive compression, where data is encoded in multiple stages to maintain high fidelity in key regions while aggressively compressing less critical data, could offer a path forward for improving 3DGS techniques, especially in complex 3D scenes where different regions (eg., background vs. foreground) require different levels of precision.

Additionally, NeRF compression methods such as \cite{tang2022compressible,deng2023compressing} have highlighted the potential of sparsity in 3D data, where significant portions of the scene can be represented with minimal data points. For 3DGS, leveraging sparse representations could lead to substantial gains in compression efficiency, as Gaussian components that are not central to the scene’s overall structure could be pruned or encoded with fewer parameters. This sparsity, combined with efficient encoding methods inspired by NeRFs, could enable next-generation 3DGS algorithms to handle large-scale 3D data more effectively.

Furthermore, techniques like adaptive quantization and neural compression employed in NeRFs open the door for utilizing machine learning models to optimize the encoding of Gaussian components. Such models can automatically adjust the precision of 3D Gaussian parameters based on the local geometric complexity and importance of the regions being encoded, or simply depth, concept explored in~\cite{deng2022depth}. By adopting similar principles, 3DGS compression could be enhanced to dynamically adjust encoding strategies based on the structure of 3D data.

\subsection{Inspiration from Point Cloud Compression}
Point clouds are among the most important and widely used 3D representations in applications such as autonomous driving, robotics, and physics simulation. The output of 3DGS is also a point cloud, making point cloud compression essential for the efficient storage and transmission of 3DGS scenes. To achieve a favorable compression ratio, it is critical to focus on lossy compression methods and address a key question: what properties of point clouds should be preserved within a limited bitrate budget?

Despite the fact that the output of 3DGS is a point cloud, there has been little focus on combining 3DGS compression with dedicated point cloud compression techniques. While the compression methods used in 3DGS result in a compressed point cloud, this is more of a byproduct of those methods rather than a direct effort to compress the point cloud itself. Given the growing importance of 3D representations in various applications, point cloud compression is a popular research area. Inspired by recent advancements in point cloud compression~\cite{he2022density,song2023efficient}, it is worth exploring how these techniques can be integrated into the end-to-end training process of 3DGS or developed as a separate module specifically tailored for compressing the output of 3DGS.

\section{Conclusion}

This survey provides a comprehensive overview of 3DGS compression methods within the framework of the proposed taxonomy. Compression techniques are categorized into unstructured and structured methods, highlighting significant advancements in memory efficiency and computational performance. We evaluate these techniques in terms of fidelity, compression ratios, and rendering speeds.  Both approaches offer distinct advantages and trade-offs. Structured compression methods leverage the relationships between Gaussians to achieve compression ratios of up to 100$\times$ compared to baseline 3DGS while maintaining high fidelity and perceptual quality. However, this comes at the cost of rendering speed, with FPS comparable to baseline 3DGS-30k. In contrast, unstructured compression methods achieve compression ratios of up to 50$\times$, preserving fidelity similar to baseline 3DGS-30k, while significantly improving rendering speeds by up to 7$\times$. These techniques address critical challenges such as storage constraints and computational overhead, further establishing 3DGS as a scalable and efficient representation for applications ranging from virtual reality to autonomous systems.  

Despite these advances, the scalability of 3DGS for large and complex scenes remains a significant challenge, especially for resource-constrained environments like mobile AR/VR devices. The field also lacks a unified framework that integrates the strengths of structured and unstructured compression methods, limiting the adaptability and standardization of 3DGS models. Furthermore, existing research has not sufficiently explored the potential of novel scalar quantization techniques or the optimization of loss functions tailored specifically for 3DGS. Addressing these gaps could unlock higher compression efficiencies, improved rendering speeds, and enhanced fidelity, making 3DGS more practical for real-world deployment. Additionally, interdisciplinary approaches combining insights from NeRFs, point-based rendering, and emerging deep learning paradigms hold promise for overcoming these limitations.

Looking forward, the future of 3DGS lies in developing hybrid compression frameworks, optimizing models for edge devices, and expanding its applications across diverse domains. Scalar quantization techniques, loss function innovations, and hardware-aware optimizations will be critical to enabling real-time rendering on low-power devices. The impact of 3DGS extends beyond academic research to industries such as healthcare, where it can enhance simulation fidelity, and entertainment, where it enables immersive experiences. By addressing current challenges, the transformative potential of 3DGS can be fully realized, redefining how machines and humans interact with 3D environments. Through continued innovation and exploration, 3DGS can become a cornerstone technology in the next generation of 3D scene rendering.

\bibliographystyle{ieeetr}
\bibliography{ref}

\end{document}

%% file: Tables/Taxonomy_Table.tex
\begin{table*}[t]
    \caption{Taxonomy and representative publications of 3DGS compression methods.}
    \label{tab:Taxonomy}
    \centering
    \footnotesize 
    \renewcommand{\arraystretch}{1.2} 
    \resizebox{0.95\textwidth}{!}{ 
    \begin{tabular}{l l p{10cm}} 
    \toprule
    \textbf{Category} & \textbf{Sub-Type} & \textbf{Representative Publications} \\
    \midrule
    \multirow{9}{*}{\textbf{Unstructured Compression}} 
        & \textbf{Pruning} 
            & LightGaussian~\cite{fan2023lightgaussian}, Compact 3D~\cite{lee2023compact}, CompGS~\cite{navaneet2023compact3d}, 
              EAGLES~\cite{girish2023eagles}, Papantonakis et al.~\cite{papantonakis2024reducing}, 
              Trimming the Fat~\cite{salman2024trimming}, SafeguardGS~\cite{lee2024safeguardgs}, EfficientGS~\cite{liu2024efficientgs}, 
              RDO-Gaussian~\cite{wang2024end}, Pup 3D-GS~\cite{hanson2024pup}, Kim et al.~\cite{kim2024color}, 
              RTGS~\cite{lin2024rtgs}, Kheradmand et al.~\cite{kheradmand20243d}, LP-3DGS~\cite{zhang2024lp}, 
              Mini-Splatting~\cite{fang2024mini}, GoDe~\cite{di2025gode}, ELMGS~\cite{ali2024elmgs} \\
        \cmidrule(l){2-3}
        & \textbf{Quantization} 
            & LightGaussian~\cite{fan2023lightgaussian}, Compact 3D~\cite{lee2023compact}, CompGS~\cite{navaneet2023compact3d}, 
              Niedermayr et al.~\cite{niedermayr2024compressed}, EAGLES~\cite{girish2023eagles}, Papantonakis et al.~\cite{papantonakis2024reducing}, 
              RDO-Gaussian~\cite{wang2024end}, Morgenstern et al.~\cite{morgenstern2023compact}, GoDe~\cite{di2025gode}, 
              ELMGS~\cite{ali2024elmgs} \\
        \cmidrule(l){2-3}
        & \textbf{Entropy Coding} 
            & Compact 3D~\cite{lee2023compact}, CompGS~\cite{navaneet2023compact3d}, Niedermayr et al.~\cite{niedermayr2024compressed}, 
              EAGLES~\cite{girish2023eagles}, Morgenstern et al.~\cite{morgenstern2023compact}, ELMGS~\cite{ali2024elmgs} \\
    \midrule
    \multirow{5}{*}{\textbf{Structured Compression}} 
        & \textbf{Graph-Based} 
            & SUNDAE~\cite{yang2024spectrally}, Gaussian-Forest (GF)~\cite{zhang2024gaussian}, Liu et al.~\cite{liu2024compgs} \\
        \cmidrule(l){2-3}
        & \textbf{Anchor-Based} 
            & Scaffold-GS~\cite{lu2023scaffold}, HAC~\cite{chen2024hac}, Context-GS~\cite{wang2024contextgs} \\
        \cmidrule(l){2-3}
        & \textbf{Contextual/AR Modeling} 
            & Context-GS~\cite{wang2024contextgs} \\
        \cmidrule(l){2-3}
        & \textbf{Factorization Approach} 
            & F-3DGS~\cite{sun2024f}, Radsplat~\cite{niemeyer2024radsplat} \\
    \bottomrule
    \end{tabular}
    }
\end{table*}

%% file: Display_Figures/taxonomy.tex
\begin{figure}[t]
    \centering
    \includegraphics[width=\linewidth]{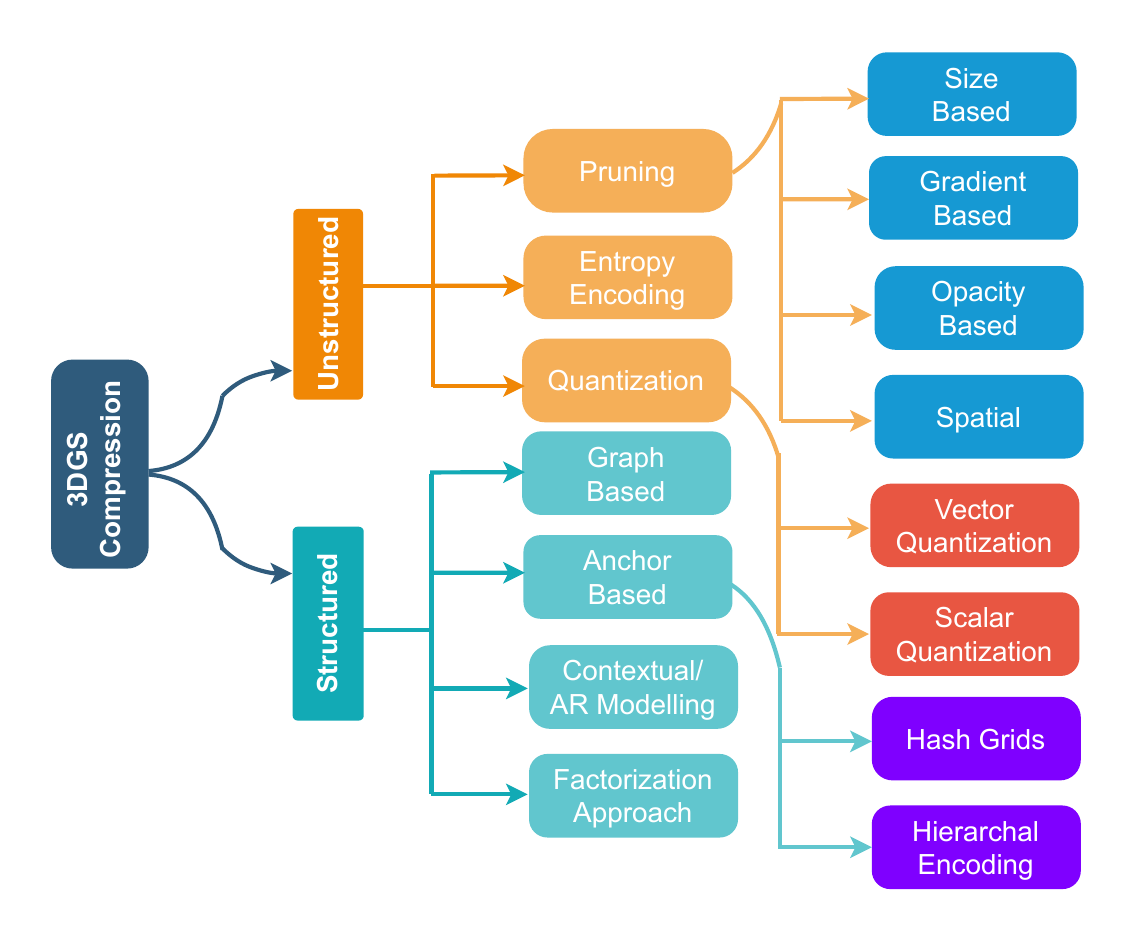}
    \caption{Taxonomy of 3DGS compression methods, detailing the structured and unstructured methods.}
    \label{fig:taxonomy}
    \vspace{-0.5cm}
\end{figure}

%% file: Display_Figures/3DGS_working.tex
\begin{figure*}[t]
    \centering
    \includegraphics[width=0.9\linewidth]{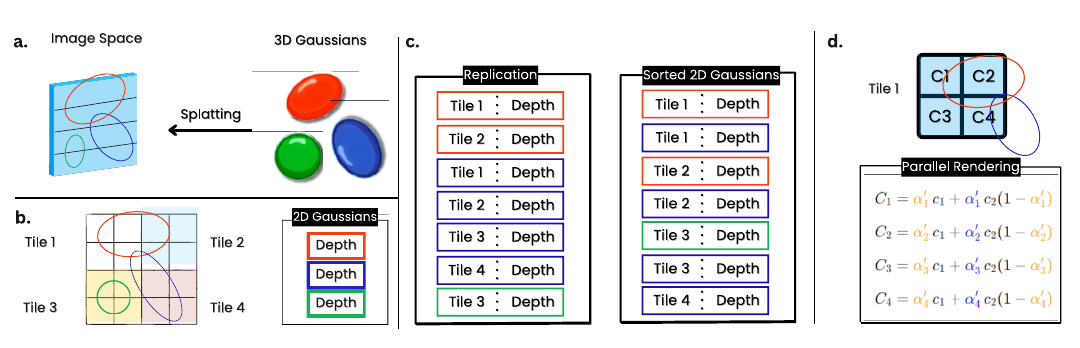}
    \caption{An overview of the 3DGS forward process. (a) The splatting step maps 3D Gaussians into the image plane. (b) 3DGS partitions the image into non-overlapping patches, referred to as tiles. (c) For Gaussians spanning multiple tiles, 3DGS duplicates them and assigns each a unique identifier, i.e., a tile ID. (d) Rendering the sorted Gaussians yields the pixel data for each tile. Note that the pixel and tile computation workflows are independent, enabling parallel processing. Best viewed in color. Figure inspired from~\cite{DBLP:journals/corr/abs-2401-03890}.}
    \label{fig:working}
\end{figure*}

%% file: Sections/Preliminaries_arxiv.tex
\section{Preliminaries}

\subsection{Differentiable Rendering }

Differentiable rendering (DR) facilitates end-to-end optimization by computing gradients of the rendering process, bridging 2D and 3D processing~\cite{kato2020differentiable}. These gradients, with respect to scene parameters like geometry, lighting, and camera pose, are crucial for tasks such as human pose estimation~\cite{ge20193d,baek2019pushing}, 3D reconstruction~\cite{yan2016perspective,tulsiani2017multi}, and neural rendering~\cite{kato2018neural}, enabling self-supervised optimization. Rendering methods are categorized as implicit or explicit. Implicit approaches represent scenes using continuous functions, such as neural networks~\cite{sitzmann2020implicit} or signed distance fields~\cite{oleynikova2016signed}, and rely on volume rendering~\cite{drebin1988volume} for image generation. Explicit methods, in contrast, define geometry using discrete primitives like triangles, points, or splats. Explicit point-based rendering focuses on realistic image synthesis by rendering collections of discrete geometric primitives. Zwicker et al.\cite{kopanas2021point} advanced this with a differentiable point-based pipeline featuring bi-directional Elliptical Weighted Average splatting, probabilistic depth testing, and efficient camera selection.

Recent advances in implicit representation techniques have led researchers to explore point-based rendering within the framework of neural implicit representations. This approach eliminates the need for predefined geometries in 3D reconstruction. A prominent example in this area is  NeRF~\cite{mildenhall2020nerf}, which uses an implicit density field to model 3D geometry and an appearance field to predict view-dependent colors. In NeRF, point-based rendering is used to aggregate the colors from all sample points along a camera ray to compute the final pixel color

\[ C = \sum_{i=1}^{N} c_i \alpha_i T_i , \]

\noindent
where \( N \) represents the number of sample points along a ray. The view-dependent color and opacity value for the \( i \)-th point on the ray are given by:

\[ \alpha_i = \exp \left(- \sum_{j=1}^{i-1} \sigma_j \delta_j \right), \]
\noindent
where \( \sigma_j \) denotes the density value of the \( j \)-th point and \( \delta_j \) is the distance between consecutive sample points. The accumulated transmittance \( T_i \) is calculated as:

\[ T_i = \prod_{j=1}^{i-1} (1 - \alpha_j). \]

The rendering process in 3DGS shares similarities with NeRFs; however, there are two key differences. \newline
\textbf{Opacity Modeling:} 3DGS directly models opacity values, while NeRF first models density values and then converts these densities into opacity values.\newline
\textbf{Rendering Technique:} 3DGS utilizes rasterization-based rendering, which avoids the need for extensive point sampling, unlike NeRF, which relies on sampling points along rays for rendering.

\subsection{3D Gaussian Splatting}
3DGS represents a significant advancement in real-time, high-resolution image rendering that does not rely on deep neural networks. 
A scene optimized with 3DGS achieves rendering by projecting the Gaussians onto a 2D image plane through splatting. After this projection, 3DGS sorts the Gaussians and computes the pixel values accordingly~\cite{kerbl20233d}. The following sections will first define the 3D Gaussian, the fundamental element of the 3DGS scene representation, then detail the differentiable rendering process used in 3DGS, and finally, describe the techniques employed to achieve high rendering speeds.

\noindent
\textbf{3D Gaussian: }The characteristics of a 3D Gaussian include its color \( c \), 3D covariance matrix \( \Sigma \), opacity \( \alpha \), and center (position) \( \mu \). For view-dependent appearance, spherical harmonics (SH) are employed to represent color \( c \). Through back-propagation, all these attributes can be learned and optimized.

\noindent
\textbf{3D Gaussian Frustum Culling: } This stage identifies the 3D Gaussians that lie outside the camera's frustum for a given camera position. Excluding these Gaussians from subsequent computations conserves processing resources, as only the 3D Gaussians within the specified view are considered.

\noindent
\textbf{Splatting of Gaussians: }In this stage, 3D Gaussians (ellipsoids) are projected onto 2D image space (ellipses) for rendering. The projected 2D covariance matrix \( \Sigma' \) is computed using the viewing transformation \( W \) and the 3D covariance matrix \( \Sigma \) as
\noindent
\[ \Sigma' = JW\Sigma W^\top J ,\]
\noindent
where \( W \) is the viewing transformation matrix, \( \Sigma \) is the 3D covariance matrix, and \( J \) is the Jacobian of the projection transformation~\cite{kerbl20233d, zwicker2001ewa}.

\noindent
\textbf{Pixel Rendering: } The viewing transformation \( W \) is employed to compute the distance between each pixel \( x \) and all overlapping Gaussians, effectively determining their depths. This results in a sorted list of Gaussians \( N \). The final color of each pixel is subsequently determined using alpha compositing. The final color \( C \) is computed by summing the contributions of each Gaussian:

\[ C = \sum_{n=1}^{|N|} c_n \alpha'_n \prod_{j=1}^{n-1} (1 - \alpha'_j) ,\]

\noindent
where \( c_n \) is the learned color. The final opacity \( \alpha'_n \) is the product of the learned opacity \( \alpha_n \) and the Gaussian function:

\[ \alpha'_n = \alpha_n \times \exp \left( -\frac{1}{2} (x' - \mu'_n)^\top \Sigma'^{-1}_n (x' - \mu'_n) \right) ,\]

\noindent
where \( x' \) is the projected position, \( \mu'_n \) is the projected center, and \( \Sigma'_n \) is the projected 2D covariance matrix. There is a valid concern that the described rendering process could be slower than NeRFs due to the challenges in parallelizing the generation of the necessary sorted list. This concern is valid, as relying on a straightforward pixel-by-pixel approach could significantly impact rendering speeds. To achieve real-time rendering, 3DGS must consider several factors that facilitate parallel processing.

\input{Display_Figures/dataset}
\input{Display_Figures/opacity_gradient_pruning_comparison}
\input{Display_Figures/opacity_distribution}
\noindent
\textbf{Tiling: }
3DGS reduces computational costs by shifting rendering accuracy from fine-grained, pixel-level detail to a more coarse, patch-level approach. The image is first divided into non-overlapping patches or tiles, and then 3DGS identifies which tiles intersect with the projected Gaussians. If a single projected Gaussian spans multiple tiles, it duplicates the Gaussian, assigning each copy a unique tile ID corresponding to the relevant tile as shown in Figure~\ref{fig:working}.

\noindent
\textbf{Rendering Parallelization: }Following the Gaussians' replication, 3DGS associates each corresponding tile ID with the depth values obtained from the view transformation. This process generates an unsorted list of bytes where the higher bits represent the tile ID and the lower bits encode the depth information. This list can then be sorted for rendering (alpha compositing). This method is particularly well-suited for parallel processing because it allows each tile and pixel to be handled independently as shown in the Figure~\ref{fig:working}. Additionally, the advantage of allowing each pixel of the tile to access shared memory while maintaining a unified read pattern enables more efficient parallel processing during the alpha compositing stage. The framework essentially treats tile and pixel processing similarly to blocks and threads in CUDA programming, as demonstrated in the original study's implementation.


In summary, 3DGS enhances computational efficiency while preserving high image reconstruction quality by incorporating several approximations during the rendering process.

\noindent
\textbf{3D Gaussian Splatting Optimization: }
The key component of 3DGS lies in an optimization process aimed at generating a substantial number of Gaussians that effectively encapsulate the key features of the scene, thereby enabling real-time 3D scene rendering. Differentiable rendering is employed to fine-tune the parameters of these 3D Gaussians to align with the scene's textures. However, the optimal number of 3D Gaussians required to accurately represent a scene is not predetermined. 
The optimization workflow encompasses a series of interrelated procedures.

\noindent
\textbf{Loss Function: }
The disparity between the ground truth and the rendered output can be evaluated upon completion of the image synthesis. To achieve this, the \( \ell_1 \) loss and D-SSIM (Differentiable Structural Similarity Index)~\cite{bakurov2022structural} loss functions are utilized. Stochastic gradient descent (SGD) is then employed to optimize all the learnable parameters. The loss function used for 3DGS is given by~\cite{kerbl20233d}:

\[ L = (1 - \lambda)L_1 + \lambda L_{\text{D-SSIM}}, \]
\noindent
where \( \lambda \in [0, 1] \) is a weighting factor.

\noindent
\textbf{Parameter Optimization:}
Back-propagation can be utilized to directly optimize most attributes of a 3D Gaussian. 
It is crucial to note that optimization of the covariance matrix directly could yield a non-positive semi-definite matrix, thereby contradicting the physical restrictions typically imposed on covariance matrices.
To address this issue, 3DGS employs an alternative approach by optimizing a quaternion \(q\) and a 3D vector \(s\), where \(q\) represents rotation and \(s\) represents scale. This method ensures that the resulting covariance matrix adheres to the required physical properties. This method enables the covariance matrix \(\Sigma\) to be reconstructed using the following formula:

\[
\Sigma = R S S^\top R^\top,
\]
\noindent
where \(S\) and \(R\) represent the scaling and rotation matrices derived from \(s\) and \(q\), respectively. The process of calculating opacity \(\alpha\) involves a complex computational graph: \(s\) and \(q \to \Sigma\), \(\Sigma \to \Sigma'\), and \(\Sigma' \to \alpha\). To avoid the computational expense of automatic differentiation, 3DGS directly derives and computes the gradients for \(q\) and \(s\) during optimization.

\noindent
\textbf{Densification Process:}
In 3DGS, optimization parameters are initialized using sparse points from Structure-from-Motion (SfM)~\cite{snavely2006photo}. Densification and pruning techniques are then applied to optimize the density of 3D Gaussians.
During densification, 3DGS adaptively increases Gaussian density in regions with sparse coverage or missing geometric features. After specific training intervals, Gaussians with gradients above a threshold are densified: large Gaussians are split into smaller ones, and small Gaussians are duplicated and shifted along gradient directions. This ensures an optimal distribution for improved 3D scene synthesis.
In the pruning stage, redundant or insignificant Gaussians are removed to regularize the representation. Gaussians with excessive view-space size or opacity \( \alpha \) below a threshold are discarded, ensuring efficiency while preserving accuracy in scene representation.

%% file: Display_Figures/dataset.tex
\begin{figure*}
    \centering
    \includegraphics[width=0.85\linewidth]{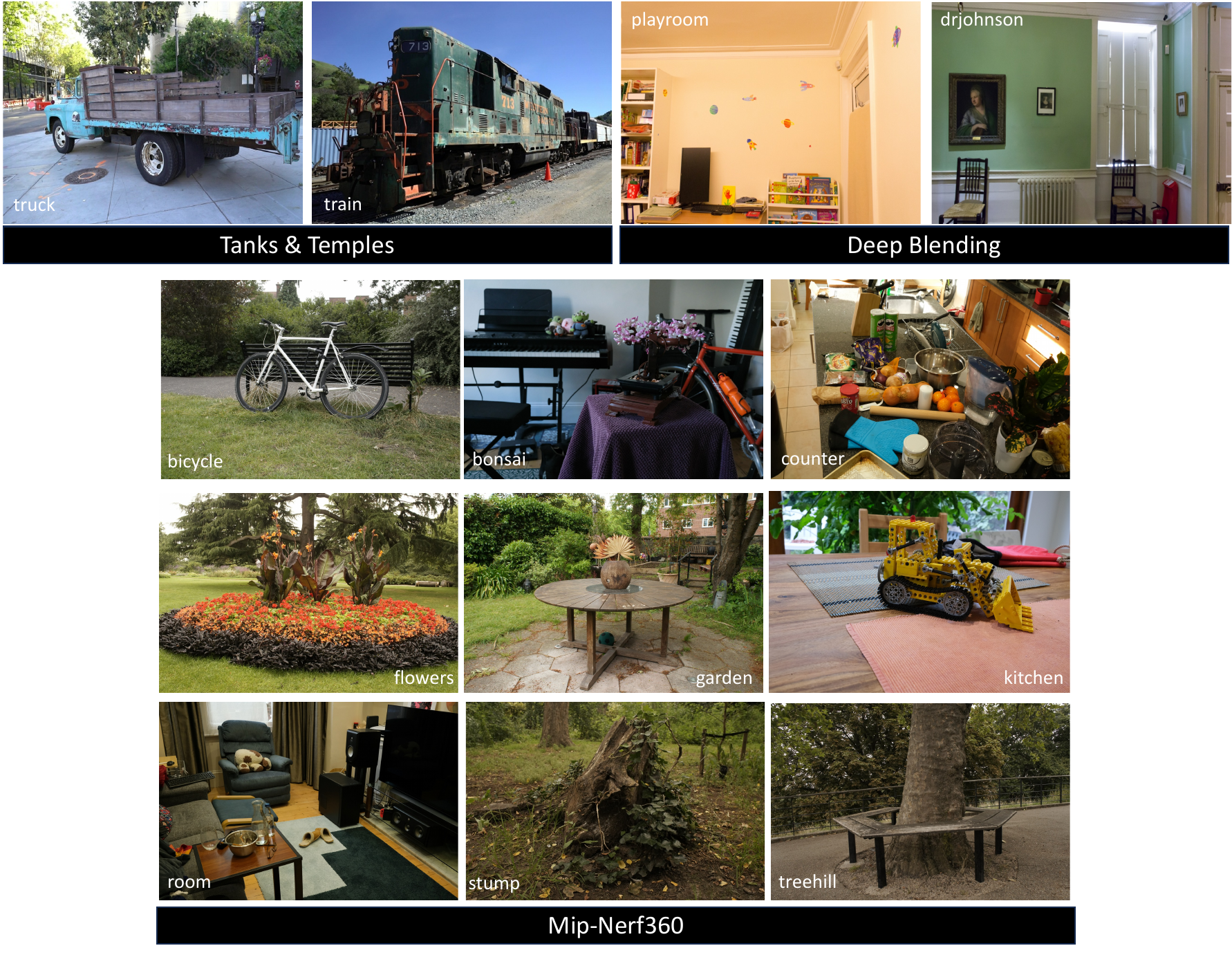}
    \caption{A representative image from each evaluation dataset showcases the diversity of scene types. The chosen images encompass small to medium-sized natural environments from Tanks\&Temples~\cite{knapitsch2017tanks}, Deep Blending~\cite{hedman2018deep}, and Mip-NeRF360~\cite{barron2022mip}, ensuring a comprehensive visual comparison across different datasets.}
    \label{fig:dataset}
    \vspace{-0.5cm}
\end{figure*}

%% file: Display_Figures/opacity_gradient_pruning_comparison.tex
\begin{figure*}[t]
    \centering
    \includegraphics[width=\linewidth]{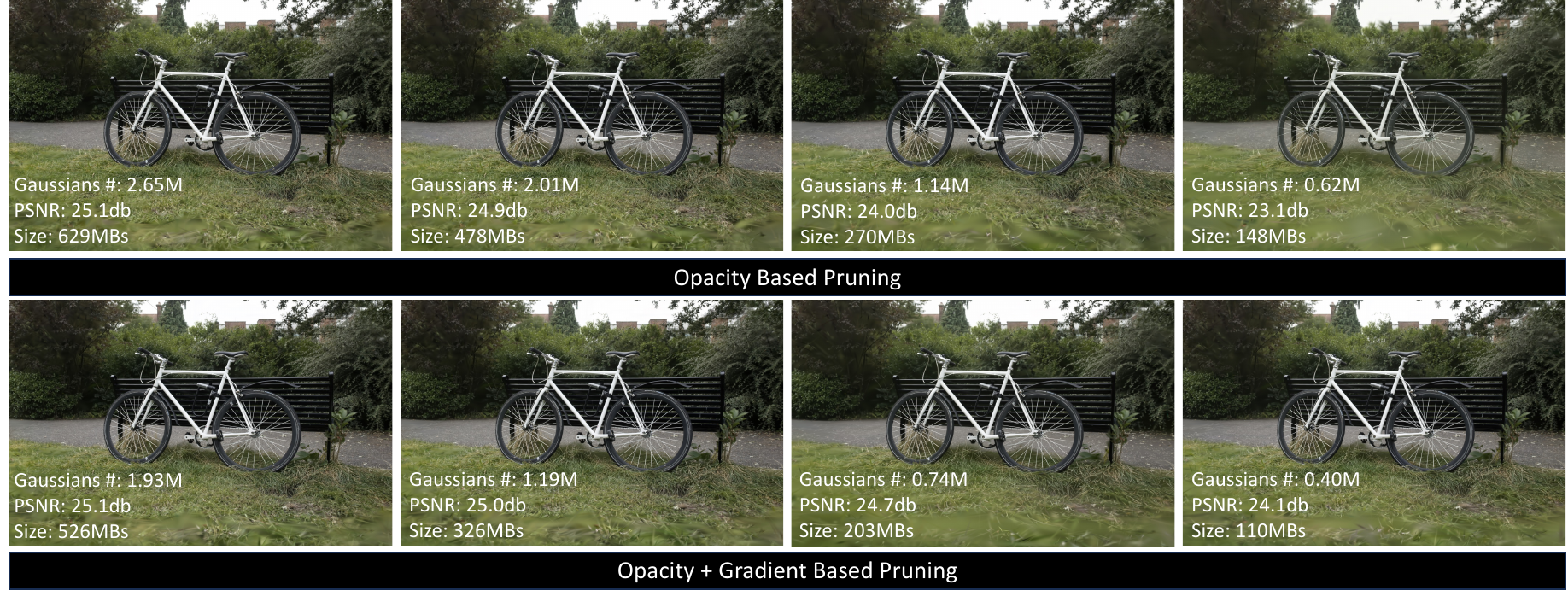}
    \caption{Comparison between opacity based pruning and gradient + opacity based pruning. The number of Gaussians are in millions (M).}
    \label{fig:opacity_gradient_pruning_comparison}
\end{figure*}

%% file: Display_Figures/opacity_distribution.tex
\begin{figure}[t]
    \centering
    \includegraphics[width=0.8\linewidth]{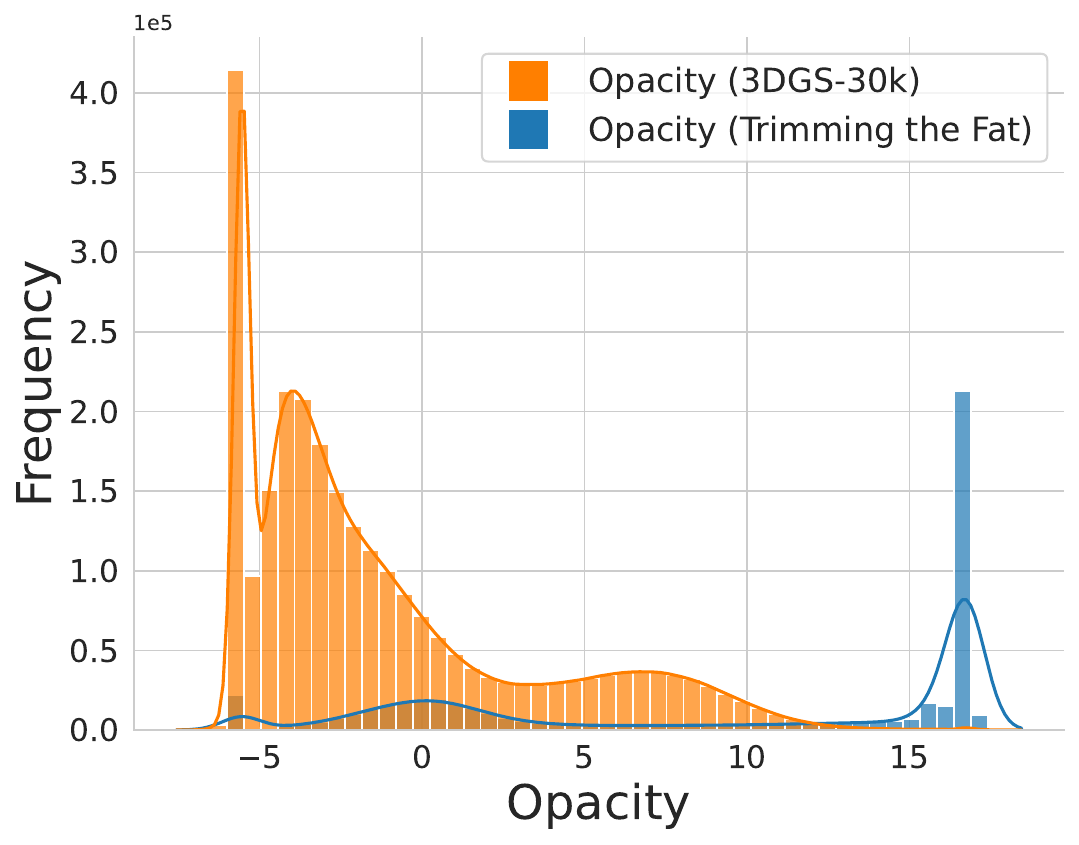}
    \caption{Opacity distribution before and after pruning for the \texttt{truck} scene. Figure taken from~\cite{salman2024trimming}.}
    \label{fig:opacity_distribution}
\end{figure}

%% file: Tables/Unstruct_table.tex
\begin{table*}[ht]
    \caption{Performance and compression comparison of the 3DGS baseline with unstructured compression methods on the Mip-NeRF360 and Tanks\&Temple datasets. All values are sourced from their respective papers. Memory size is reported in megabytes (MB). The best-performing results are highlighted in red, followed by yellow and green.}
    \label{tab:unstruc_mip_tandt}
    \centering
    \small
    \resizebox{0.95\textwidth}{!}{
    \begin{tabular}{l|ccccc|ccccc}
    \toprule
    &  \multicolumn{5}{c}{\textbf{Mip-NeRF360~\cite{barron2022mip}}} & \multicolumn{5}{c}{\textbf{Tanks\&Temples~\cite{knapitsch2017tanks}}} \\
    \textbf{Model} & \textbf{SSIM$^\uparrow$} & \textbf{PSNR$^\uparrow$} & \textbf{LPIPS$^\downarrow$} & \textbf{Mem.}$^\downarrow$ & \textbf{Comp.$^\uparrow$} & \textbf{SSIM$^\uparrow$} & \textbf{PSNR$^\uparrow$} & \textbf{LPIPS$^\downarrow$} & \textbf{Mem.}$^\downarrow$ & \textbf{Comp.$^\uparrow$}\\
    \midrule

3DGS-30k~\cite{kerbl20233d} & \cellcolor{yellow!25}0.815 & 27.21 & \cellcolor{red!25}0.214 & 734.0 & 1$\times$ & \cellcolor{red!25}0.841 & 23.14 & \cellcolor{red!25}0.183 & 411.0 & 1$\times$ \\
\midrule
LightGaussian~\cite{fan2023lightgaussian} & 0.805 & \cellcolor{green!25}27.28 & 0.243 & 42.0 & 18$\times$ & 0.817 & 23.11 & 0.231 & 22.0 & 19$\times$ \\
Compact3D~\cite{lee2023compact} & 0.798 & 27.08 & 0.247 & 48.8 & 15$\times$ & 0.831 & 23.32 & 0.201 & 39.4 & 10$\times$ \\
CompGS-16K~\cite{navaneet2023compact3d} & 0.804 & 27.03 & 0.243 & \cellcolor{red!25}18.0 & \cellcolor{red!25}41$\times$ & 0.836 & 23.39 & 0.200 & \cellcolor{green!25}12.0 & \cellcolor{green!25}34$\times$ \\
CompGS-32K~\cite{navaneet2023compact3d} & 0.806 & 27.12 & 0.240 & \cellcolor{yellow!25}19.0 & \cellcolor{yellow!25}39$\times$ & \cellcolor{green!25}0.838 & 23.44 & 0.198 & 13.0 & 32$\times$ \\
Niedermayr et al.~\cite{niedermayr2024compressed} & 0.801 & 26.98 & 0.238 & 28.8 & 26$\times$ & 0.832 & 23.32 & \cellcolor{green!25}0.194 & 17.3 & 24$\times$ \\
EAGLES~\cite{girish2023eagles} & \cellcolor{green!25}0.810 & 27.23 & 0.240 & 54.0 & 14$\times$ & \cellcolor{yellow!25}0.840 & 23.37 & 0.200 & 29.0 & 14$\times$ \\
Papantonakis et al.~\cite{papantonakis2024reducing} & 0.809 & 27.10 & \cellcolor{green!25}0.226 & 29.0 & 25$\times$ & \cellcolor{yellow!25}0.840 & 23.57 & \cellcolor{yellow!25}0.188 & 14.0 & 29$\times$ \\
Trimming the Fat~\cite{salman2024trimming} & 0.798 & 27.13 & 0.248 & \cellcolor{green!25}20.1 & \cellcolor{green!25}37$\times$ & 0.831 & \cellcolor{green!25}23.69 & 0.210 & \cellcolor{red!25}8.6 & \cellcolor{red!25}48$\times$ \\
Efficientgs~\cite{liu2024efficientgs} & \cellcolor{red!25}0.817 & \cellcolor{red!25}27.38 & \cellcolor{yellow!25}0.216 & 98.0 & 8$\times$ & 0.837 & 23.45 & 0.197 & 33.0 & 13$\times$ \\
RDO-Gaussian~\cite{wang2024end} & 0.802 & 27.05 & 0.239 & 23.5 & 31$\times$ & 0.835 & 23.34 & 0.195 & \cellcolor{green!25}12.0 & \cellcolor{green!25}34$\times$ \\
PUP 3D-GS~\cite{hanson2024pup} & 0.792 & 26.83 & 0.268 & 86.3 & 9$\times$ & 0.807 & 23.03 & 0.245 & 50.1 & 8$\times$ \\
Kim et al.~\cite{kim2024color} & 0.797 & 27.07 & 0.249 & 73.0 & 10$\times$ & 0.830 & 23.18 & 0.198 & 42.0 & 10$\times$ \\
ELMGS-medium~\cite{ali2024elmgs} & 0.792 & \cellcolor{yellow!25}27.31 & 0.264 & 38.6 & 19$\times$ & 0.838 & \cellcolor{red!25}24.08 & 0.191 & 18.8 & 22$\times$\\
ELMGS-small~\cite{ali2024elmgs} & 0.779 & 27.00 & 0.286 & 25.8 & 28$\times$ & 0.825 & \cellcolor{yellow!25}23.90 & 0.233 & \cellcolor{yellow!25}11.6 & \cellcolor{yellow!25}35$\times$\\

    \bottomrule
    \end{tabular}
    }
\end{table*}

\begin{table}[ht]
    \caption{Performance and compression comparison of the 3DGS baseline with unstructured compression methods on the Deep Blending dataset. All values are sourced from their respective papers. Memory size is reported in megabytes (MB). The best-performing results are highlighted in red, followed by yellow and green.}
    \label{tab:unstruc_db}
    \centering
    \small
    \resizebox{0.49\textwidth}{!}{

    \begin{tabular}{l|ccccc}
    \toprule
    &  \multicolumn{4}{c}{\textbf{Deep Blending~\cite{hedman2018deep}}}  \\
    \textbf{Model} & \textbf{SSIM$^\uparrow$} & \textbf{PSNR$^\uparrow$} & \textbf{LPIPS$^\downarrow$} & \textbf{Mem.}$^\downarrow$ & \textbf{Comp.$^\uparrow$} \\
    \midrule
3DGS-30k~\cite{kerbl20233d} & 0.903 & 29.41 & \cellcolor{red!25}0.243 & 676.0 & 1$\times$ \\
\midrule
Compact3D~\cite{lee2023compact} & 0.901 & \cellcolor{green!25}29.79 & 0.258 & 43.2 & 16$\times$ \\
CompGS-16K~\cite{navaneet2023compact3d} & \cellcolor{green!25}0.906 & \cellcolor{red!25}29.90 & 0.252 & \cellcolor{red!25}12.0 & \cellcolor{red!25}56$\times$ \\
CompGS-32K~\cite{navaneet2023compact3d} & \cellcolor{yellow!25}0.907 & \cellcolor{red!25}29.90 & 0.251 & 13.0 & 52$\times$ \\
Niedermayr et al.~\cite{niedermayr2024compressed} & 0.898 & 29.38 & 0.253 & 25.3 & 27$\times$ \\
EAGLES~\cite{girish2023eagles} & \cellcolor{red!25}0.910 & \cellcolor{yellow!25}29.86 & \cellcolor{green!25}0.250 & 52.0 & 13$\times$ \\
Papantonakis et al.~\cite{papantonakis2024reducing} & 0.902 & 29.63 & \cellcolor{yellow!25}0.249 & 18.0 & 38$\times$ \\
Trimming the Fat~\cite{salman2024trimming} & 0.897 & 29.43 & 0.267 & \cellcolor{green!25}12.5 & \cellcolor{green!25}54$\times$ \\
Efficientgs~\cite{liu2024efficientgs} & 0.903 & 29.63 & 0.251 & 40.0 & 17$\times$ \\
RDO-Gaussian~\cite{wang2024end} & 0.902 & 29.63 & 0.252 & 18.0 & 38$\times$ \\
PUP 3D-GS~\cite{hanson2024pup} & 0.881 & 28.61 & 0.305 & 80.8 & 8$\times$ \\
Kim et al.~\cite{kim2024color} & 0.902 & 29.71 & 0.255 & 72.0 & 9$\times$ \\
ELMGS-medium~\cite{ali2024elmgs} & 0.897 & 29.48 & 0.261 & 23.5 & 29$\times$\\
ELMGS-small~\cite{ali2024elmgs} & 0.894 & 29.24 & 0.273 & \cellcolor{yellow!25}12.3 & \cellcolor{yellow!25}55$\times$\\
    \bottomrule
    \end{tabular}
}
\vspace{-0.5cm}
\end{table}

%% file: Display_Figures/unstruct_fps_vs_psnr.tex
\begin{figure*}[t]
    \centering
    \includegraphics[width=\linewidth]{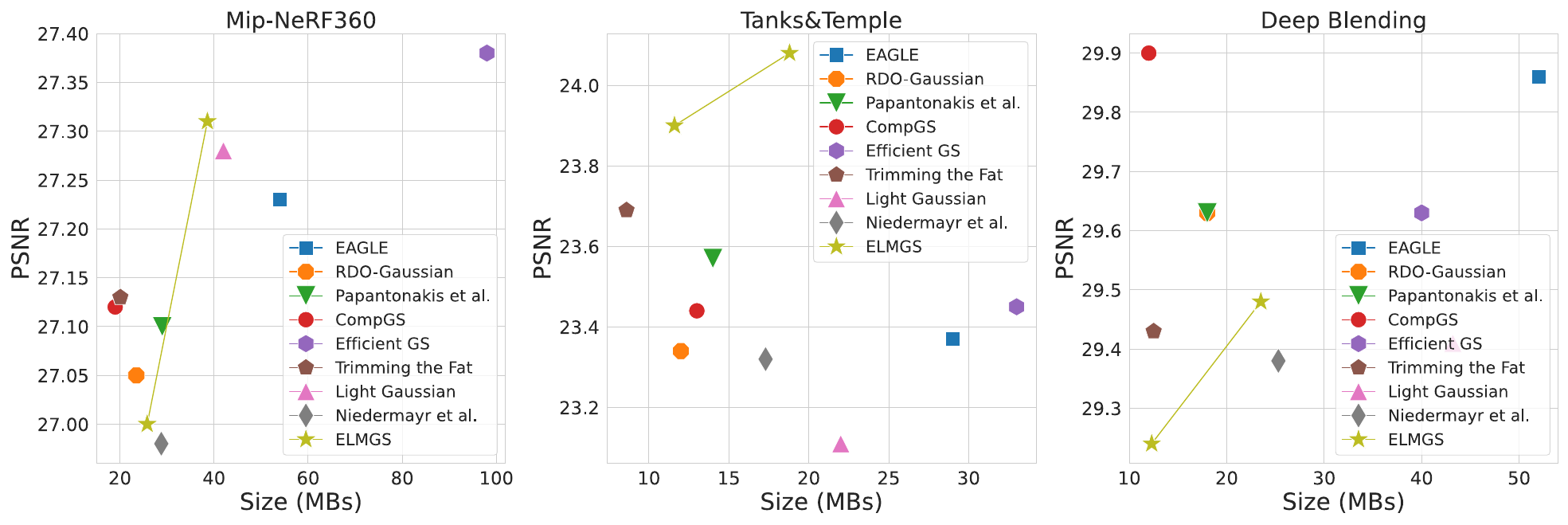}
    \caption{Comparison of PSNR and storage size across unstructured compression techniques for the Mip-NeRF360, Tanks \& Temple, and Deep Blending datasets.}
    \label{fig:unstruc_memvspsnr}
    \vspace{-0.5cm}
\end{figure*}

%% file: Display_Figures/unstruct_mem_vs_psnr.tex
\begin{figure*}[t]
    \centering
    \includegraphics[width=\linewidth]{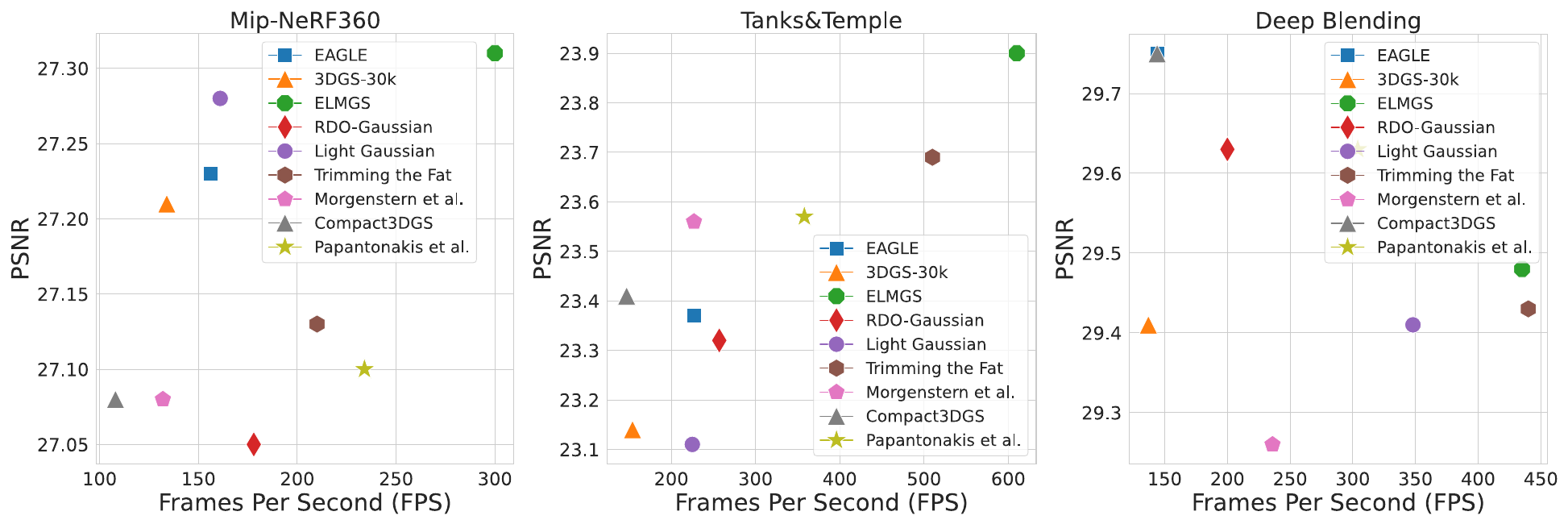}
    \caption{Comparison of PSNR and FPS across unstructured compression techniques. The FPS for all the techniques is evaluated on a single NVIDIA A40 GPU.}
    \label{fig:unstruc_fpsvspsnr}
    \vspace{-0.5cm}    
\end{figure*}

%% file: Sections/Unstructured_Compression.tex
\subsection{Discussion}
We analyze the performance of various unstructured compression techniques for 3DGS across the Mip-NeRF360, Tanks\&Temples, and Deep Blending datasets, focusing on three key aspects: compression ratio, fidelity, and FPS.

\noindent
\textbf{Fidelity: }EfficientGS~\cite{liu2024efficientgs}, Papantonakis et al.~\cite{papantonakis2024reducing}, and EAGLES~\cite{girish2023eagles} across all datasets exhibit strong performance, maintaining high SSIM, PSNR, and low LPIPS scores, indicating minimal perceptual degradation as shown in Tables~\ref{tab:unstruc_mip_tandt} and~\ref{tab:unstruc_db}. ELMGS-medium~\cite{ali2024elmgs} provides a compelling balance between compression and quality, notably achieving 24.08dB in PSNR on Tanks\&Temples outperforming all the other methods. On the other hand, highly compressed techniques such as Trimming the Fat~\cite{salman2024trimming}, CompGS~\cite{navaneet2023compact3d}, and ELMGS-small~\cite{ali2024elmgs} exhibit slightly elevated LPIPS scores, suggesting perceptual quality loss despite their strong numerical PSNR performance. These results emphasize that while EfficientGS,  Papantonakis et al~\cite{papantonakis2024reducing}, and EAGLES provide the best fidelity, ELMGS and Trimming the Fat offer promising trade-offs between compression efficiency and reconstruction quality.

\input{Display_Figures/scaffolgs}

\noindent
\textbf{Compression Ratio: } 
The results in Tables~\ref{tab:unstruc_mip_tandt} and~\ref{tab:unstruc_db} reveal significant variability in compression performance across methods and Figure~\ref{fig:unstruc_memvspsnr} shows rate-distortion tradeoff. ELMGS-small demonstrates the highest compression rates, achieving a 55$\times$ reduction on the Deep Blending dataset and 35$\times$  on Tanks\&Temples, making it one of the most storage-efficient methods. Similarly, CompGS-16K and CompGS-32K provide compression factors of 39$\times$ –56$\times$ , albeit at a minor fidelity trade-off. Notably, Trimming the Fat achieves a 48$\times$  compression on Tanks\&Temples while maintaining a competitive balance between storage complexity and reconstruction quality. 

\noindent
\textbf{FPS: }Frame rate is a crucial factor in real-time rendering applications, ensuring whether a method is viable for interactive environments such as virtual reality and gaming or low-power edge devices. The rendering speeds are calculated on a single NVIDIA A40 GPU. The FPS vs. PSNR plots in Figure~\ref{fig:unstruc_fpsvspsnr} highlight a clear trade-off between rendering speed and reconstruction accuracy. ELMGS, Trimming the Fat, and Papantonakis et al.~\cite{papantonakis2024reducing} achieve the highest FPS values, with ELMGS surpassing 400 FPS on Deep Blending and 600 FPS on Tanks\&Temples, making it particularly well-suited for real-time applications with limited computational complexity. Compared to the 3DGS-30k baseline all the unstructured compression methods offer a better trade-off between rendering speed and fidelity, making them strong candidates for real-world applications.

EfficientGS on Mip-NeRF360 and EAGLES on deep blending provide the highest fidelity. However, ELMGS offers the most aggressive compression while retaining competitive reconstruction quality. CompGS and Trimming the Fat achieve optimal trade-offs between compression, fidelity, and real-time rendering speed. These findings indicate that no single method dominates all aspects, highlighting the importance of application-specific selection when choosing an appropriate unstructured compression technique for 3DGS.

%% file: Display_Figures/scaffolgs.tex
\begin{figure*}
    \centering
    \includegraphics[width=0.9\linewidth]{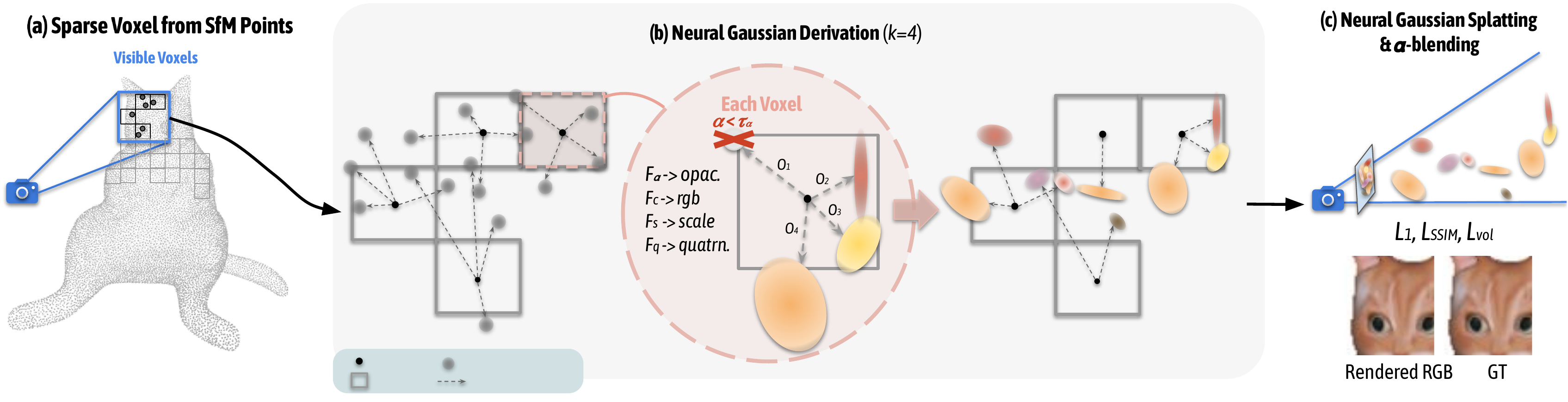}
    \caption{\textbf{Overview of the ScaffoldGS pipeline.} (a) A sparse voxel grid is initialized from Structure-from-Motion (SfM)-derived points, with each \textbf{anchor} placed at the voxel center and assigned a learnable scale, capturing the scene’s occupancy. (b) Within the view frustum, \textbf{k Gaussians} are generated from each \textit{visible anchor} with offsets \(\{O_k\}\), and their attributes—including opacity, color, scale, and quaternion—are decoded from the anchor feature. (c) To enhance efficiency and reduce redundancy, only significant Gaussians (\(\alpha \geq \tau_\alpha\)) are rasterized. The final rendered image is optimized using reconstruction loss. Figure taken from~\cite{lu2023scaffold}.}
    \label{fig:scaffoldgs}
    \vspace{-0.5cm}
\end{figure*}

%% file: Tables/struct_table.tex
\begin{table*}[]
    \caption{Performance and compression comparison of the 3DGS baseline with structured compression methods on the Mip-NeRF360 and Tanks\&Temples datasets. All values are sourced from their respective papers. Memory size is reported in megabytes (MB). The best-performing results are highlighted in red, followed by yellow and green.}
    \label{tab:structured_mip_tandt}
    \centering
    \small
    \resizebox{0.95\textwidth}{!}{
    \begin{tabular}{l|ccccc|ccccc}
    \toprule
    &  \multicolumn{5}{c}{\textbf{Mip-NeRF360~\cite{barron2022mip}}} & \multicolumn{5}{c}{\textbf{Tanks\&Temples~\cite{knapitsch2017tanks}}} \\
    \textbf{Model} & \textbf{SSIM$^\uparrow$} & \textbf{PSNR$^\uparrow$} & \textbf{LPIPS$^\downarrow$} & \textbf{Mem.}$^\downarrow$ & \textbf{Comp.$^\uparrow$} & \textbf{SSIM$^\uparrow$} & \textbf{PSNR$^\uparrow$} & \textbf{LPIPS$^\downarrow$} & \textbf{Mem.}$^\downarrow$ & \textbf{Comp.$^\uparrow$} \\
    \midrule
    
3DGS-30k~\cite{kerbl20233d} & 0.815 & 27.21 & \cellcolor{yellow!25}0.214 & 734.0 & 1$\times$ & 0.841 & 23.14 & \cellcolor{green!25}0.183 & 411.0 & 1$\times$\\
\midrule
ScaffoldGS~\cite{lu2023scaffold}            & \cellcolor{red!25}0.848 & \cellcolor{red!25}28.84 & 0.220 & 102.0  & 7$\times$ & \cellcolor{yellow!25}0.853 & 23.96 & \cellcolor{yellow!25}0.177 & 87.0  & 5$\times$ \\
GF\_Large~\cite{zhang2024gaussian}   & 0.803 & 27.45 & \cellcolor{red!25}0.212 & 85.0 & 9$\times$ & 0.839 & 23.67 & 0.188 & 45.0 & 9$\times$ \\ 
GF\_Small~\cite{zhang2024gaussian}   & 0.797 & 27.33 & \cellcolor{green!25}0.219 & 50.0 & 15$\times$ & 0.836 & 23.56 & 0.194 & 38.0 & 11$\times$ \\
SUNDAE (30\%)~\cite{yang2024spectrally}            & \cellcolor{yellow!25} 0.826 & 27.24 & 0.228 & 279.0  & 3$\times$ & 0.817 & 23.46 & 0.242 & 148.0  & 3$\times$ \\
SUNDAE (1\%)~\cite{yang2024spectrally}             & 0.716 & 24.70 & 0.375 & 38.0   & 19$\times$ & 0.703 & 20.49 & 0.375 & 33.0   & 13$\times$ \\
HAC-lowrate~\cite{chen2024hac}             & 0.807 & 27.53 & 0.238 & \cellcolor{green!25}15.3  & \cellcolor{green!25}48$\times$ & 0.846 & \cellcolor{green!25}24.04 & 0.187 & \cellcolor{green!25}8.1  & \cellcolor{green!25}51$\times$ \\
HAC-highrate~\cite{chen2024hac}            & \cellcolor{green!25}0.811 & \cellcolor{yellow!25}27.77 & 0.230 & 21.9  & 34$\times$ & \cellcolor{yellow!25}0.853 & \cellcolor{red!25}24.40 & \cellcolor{yellow!25}0.177 & 11.2 & 37$\times$ \\
ContextGS (lowrate)~\cite{wang2024contextgs}     & 0.808 & 27.62 & 0.237 & \cellcolor{yellow!25}12.7  & \cellcolor{yellow!25}58$\times$ & \cellcolor{green!25}0.852 & 24.20 & 0.184 & \cellcolor{yellow!25}7.1  & \cellcolor{yellow!25}58$\times$ \\
ContextGS (highrate)~\cite{wang2024contextgs}    & \cellcolor{green!25}0.811 & \cellcolor{green!25}27.75 & 0.231 & 18.4 & 40$\times$ & \cellcolor{red!25}0.855 & \cellcolor{yellow!25}24.29 & \cellcolor{red!25}0.176 & 11.8  & 35$\times$ \\
CompGS (highrate)~\cite{liu2024compgs}       & 0.800 & 27.26 & 0.240  & 16.5  & 45$\times$ & 0.840 & 23.70 & 0.210  & 9.6   & 43$\times$ \\
CompGS (lowrate)~\cite{liu2024compgs}        & 0.780 & 26.37 & 0.280  & \cellcolor{red!25}8.8   & \cellcolor{red!25}83$\times$ & 0.810 & 23.11 & 0.240  & \cellcolor{red!25}5.9   & \cellcolor{red!25}70$\times$ \\

    \bottomrule
    \end{tabular} 
    }
    \vspace{-0.2cm}
\end{table*}

\begin{table}[ht]
    \caption{Performance and compression comparison of the 3DGS baseline with structured compression methods on the Deep Blending dataset. All values are sourced from their respective papers. Memory size is reported in megabytes (MB). The best-performing results are highlighted in red, followed by yellow and green.}
    \label{tab:struct_db}
    \centering
    \small
    \resizebox{0.49\textwidth}{!}{

    \begin{tabular}{l|ccccc}
    \toprule
    &  \multicolumn{5}{c}{\textbf{Deep Blending~\cite{hedman2018deep}}}  \\
    \textbf{Model} & \textbf{SSIM$^\uparrow$} & \textbf{PSNR$^\uparrow$} & \textbf{LPIPS$^\downarrow$} & \textbf{Mem.}$^\downarrow$ & \textbf{Comp.$^\uparrow$} \\
    \midrule
3DGS-30k~\cite{kerbl20233d} & 0.903 & 29.41 & \cellcolor{green!25}0.243 & 676.0 & 1$\times$  \\
\midrule
ScaffoldGS~\cite{lu2023scaffold} & 0.906 & \cellcolor{green!25}30.21 & 0.254 & 66.0 & 10$\times$  \\
GF\_Large~\cite{zhang2024gaussian} & \cellcolor{yellow!25}0.908 & 30.18 & \cellcolor{red!25}0.215 & 98.0 & 7$\times$  \\
GF\_Small~\cite{zhang2024gaussian} & 0.905 & 30.11 & \cellcolor{yellow!25}0.223 & 64.0 & 11$\times$  \\
SUNDAE (30\%)~\cite{yang2024spectrally} & 0.899 & 29.40 & 0.248 & 203.0 & 3$\times$  \\
SUNDAE (1\%)~\cite{yang2024spectrally} & 0.861 & 26.57 & 0.355 & 36.0 & 19$\times$  \\
HAC-lowrate~\cite{chen2024hac} & 0.902 & 29.98 & 0.269 & \cellcolor{yellow!25}4.4 & \cellcolor{yellow!25}154$\times$  \\
HAC-highrate~\cite{chen2024hac} & 0.906 & \cellcolor{yellow!25}30.34 & 0.258 & 6.4 & 106$\times$  \\
ContextGS (lowrate)~\cite{wang2024contextgs} & \cellcolor{green!25}0.907 & 30.11 & 0.265 & \cellcolor{red!25}3.6 & \cellcolor{red!25}188$\times$  \\
ContextGS (highrate)~\cite{wang2024contextgs} & \cellcolor{red!25}0.909 & \cellcolor{red!25}30.39 & 0.258 & 6.6 & 102$\times$  \\
CompGS (highrate)~\cite{liu2024compgs}  & 0.900 & 29.69 & 0.280 & 8.8 & 77$\times$  \\
CompGS (lowrate)~\cite{liu2024compgs}  & 0.900 & 29.30 & 0.290 & \cellcolor{green!25}6.0 & \cellcolor{green!25}113$\times$  \\

    \bottomrule
    \end{tabular}
}
\vspace{-0.6cm}
\end{table}

%% file: Sections/Structured_Compression.tex
\section{Structured Compression}
Structured compression techniques for 3DGS introduce organization into the otherwise sparse Gaussian representations, enabling more efficient storage and processing. These approaches leverage spatial relationships between Gaussians through anchors, graphs, trees, hierarchical structures, and predictive models. Unlike unstructured methods, which primarily focus on pruning, quantization, and entropy encoding without incorporating the underlying structure of Gaussians, structured approaches introduce architectural constraints to optimize compression efficiency while maintaining reconstruction quality.

\noindent
\subsection{Anchor Based}
Scaffold-GS~\cite{lu2023scaffold} introduces anchor points to structure Gaussians within a scene, ensuring more compact representations while maintaining fidelity (Figure~\ref{fig:scaffoldgs}). Anchors serve as structured reference points, guiding the hierarchical organization and spatial distribution of Gaussians. Unlike traditional 3DGS, where Gaussians freely drift or split, Scaffold-GS employs a grid of anchors, initially derived from SfM points, to provide a region-aware and constrained representation. Each anchor acts as a fixed control center that tethers multiple Gaussians with learnable offsets, enabling local adaptations while preserving structural coherence. Scaffold-GS dynamically predicts Gaussian attributes such as opacity, color, rotation, and scale based on local feature representations and viewing positions. It also employs anchor growing and pruning strategies to refine scene representations—introducing new anchors in underrepresented regions and removing redundant ones. As a foundational method, Scaffold-GS serves as a backbone for subsequent structured compression techniques, providing a reliable framework for further optimization.
HAC (Hash-Grid Assisted Context Modeling)~\cite{chen2024hac}  builds upon Scaffold-GS by incorporating hash grid features to capture spatial dependencies among Gaussians. A hash grid is a structured data representation to efficiently store and query spatial data. HAC uses a binary hash grid to capture spatial consistencies among unorganized 3D Gaussians (or anchors). HAC introduces an Adaptive Scalar Quantization Module (AQM) that dynamically adjusts quantization step sizes for different anchor attributes. HAC queries the hash grid by anchor location to obtain interpolated hash features, which are then used to predict the distribution of anchor attribute values, enabling efficient entropy coding, such as Arithmetic Coding (AE)~\cite{witten1987arithmetic}, for a highly compact model representation. However, HAC encodes all anchors simultaneously, leaving room for further optimization to reduce spatial redundancy.

\noindent
\subsection{Contextual/ Autoregressive (AR) Modeling}
Context models~\cite{minnen2018joint}, commonly used in image compression (LIC) to enhance coding efficiency by predicting the distribution of latent pixels based on already coded ones, inspired ContextGS~\cite{wang2024contextgs}. This approach encodes anchor features autoregressively, predicting anchor points from those already coded at coarser levels. However, compared to LIC methods, the AR-based approach used in ContextGS is significantly faster. ContextGS employs a three-level hierarchical encoding scheme for anchors, where higher-level anchors rely on context from coarser-level anchors for both encoding and decoding. This approach closely mirrors the ChARM~\cite{minnen2020channel} method, making the process more efficient while utilizing the context for compression. Another approach CompGS (Compressed Gaussian Splatting)~\cite{li2017efficient} also utilizes context by introducing a hybrid primitive structure, where sparse anchor primitives serve as reference points for predicting attributes of other Gaussians. 

\noindent
\subsection{Graph Based}
Unlike anchor-based methods, graph-based compression treats Gaussians as a graph, leveraging their spatial relationships to optimize compression. Spectrally Pruned Gaussian Fields with Neural Compensation (SUNDAE)~\cite{yang2024spectrally} prunes redundant Gaussians while preserving key scene details. A distinct feature of SUNDAE is its neural compensation head, which shifts from direct GS to feature splatting, enabling lightweight neural network-based feature interpolation. 
GaussianForest (GF)~\cite{zhang2024gaussian} introduces a tree-structured hierarchical representation for 3D scenes. Unlike traditional unstructured Gaussians, GF assigns explicit attributes (e.g., position, opacity) to leaf nodes, while implicit attributes (e.g., covariance matrix, view-dependent color) are shared across hierarchical levels. Additionally, GF employs adaptive growth and pruning, dynamically adjusting the representation to scene complexities.

\noindent
\subsection{Factorization Approach}
Factorization-based compression techniques introduce mathematical approximations to efficiently represent dense clusters of Gaussians. F-3DGS (Factorized 3D Gaussian Splatting)~\cite{sun2024f} employs matrix and tensor factorization to represent high-dimensional Gaussian parameters using a limited number of basis components per axis. Instead of storing all parameters directly, F-3DGS factorizes attributes such as color, scale, and rotation substantially reducing storage requirements while preserving essential visual attributes. 

\subsection{Discussion}
We also analyze the performance of structured compression techniques for 3DGS on the Mip-NeRF360, Tanks\&Temples, and Deep Blending datasets, focusing on fidelity, compression ratio, and FPS.

\begin{figure*}[t]
    \centering
    \includegraphics[width=\linewidth]{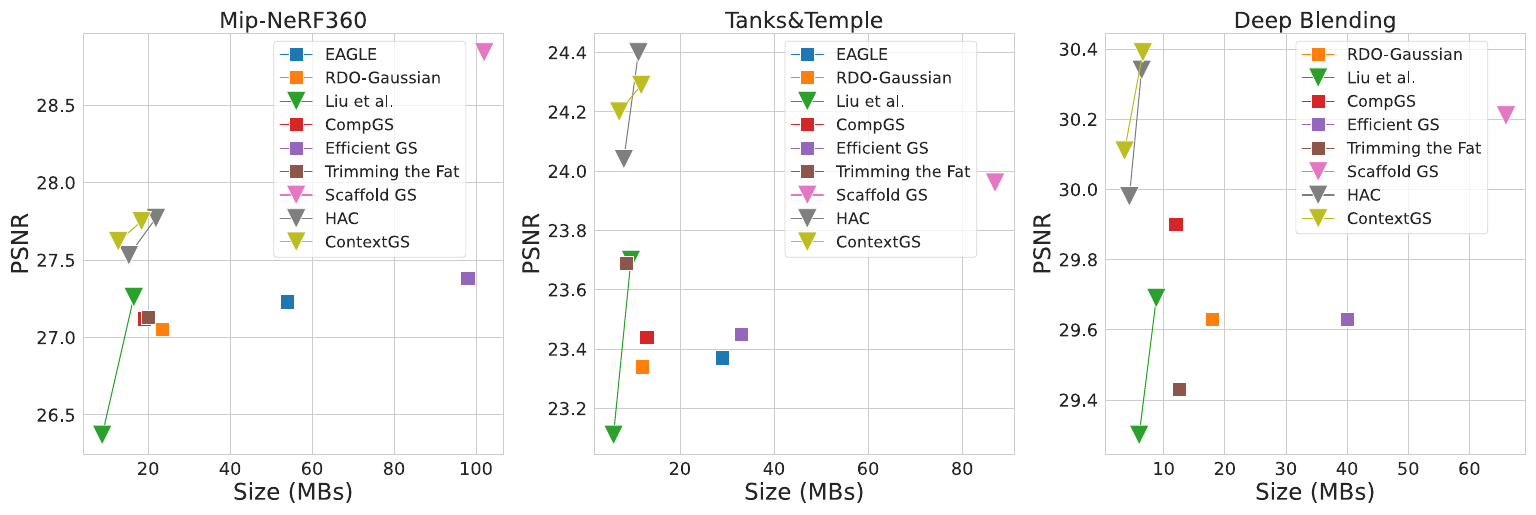}
    \caption{Comparison of PSNR and size across structured and unstructured compression techniques. The box represents unstructured compression methods, while the downward triangle represents structured compression methods.}
    \label{fig:memvspsnr}
\end{figure*}

\noindent
\textbf{Fidelity: }
Structured compression methods such as ScaffoldGS, HAC-highrate, and ContextGS (highrate) exhibit strong fidelity across all datasets as shown in Table~\ref{tab:structured_mip_tandt} and Table~\ref{tab:struct_db}. ContextGS-highrate achieves the highest PSNR of 30.39 dB and SSIM of 0.909 on Deep Blending, while HAC-highrate attains a PSNR of 30.34 dB and SSIM of 0.906. ScaffoldGS maintains a competitive PSNR of 30.21 dB and SSIM of 0.906 on Mip-NeRF360.
Interestingly, CompGS-highrate, despite achieving high PSNR, exhibits slightly higher LPIPS score, indicating some perceptual degradation. These results suggest that ScaffoldGS, ContextGS, and HAC-highrate offer the best fidelity, while CompGS provides an optimal balance between compression efficiency and perceptual quality.

\noindent
\textbf{Compression Ratio: }Tables~\ref{tab:structured_mip_tandt} and~\ref{tab:struct_db} show CompGS (lowrate) achieves the highest compression ratio, reducing memory by 83$\times$ on Mip-NeRF360, 70$\times$ on Tanks\&Temples, and 113$\times$ on Deep Blending, while ContextGS (lowrate) reaches 58$\times$ on Mip-NeRF360 and Tanks\&Temples and 188$\times$ on Deep Blending, making them the most storage-efficient techniques with lowest memory footprint compared to the baseline. HAC-lowrate also demonstrates strong compression performance, achieving 48$\times$ reduction on Mip-NeRF360 and 154$\times$ on Deep Blending, with a favorable balance between storage efficiency and fidelity. In contrast, ScaffoldGS and GF, though still reducing memory footprint, operate at relatively lower compression rates (7$\times$–10$\times$).

\noindent
\textbf{FPS: }Structured compression methods, while achieving high compression rates, often maintain FPS similar to or lower than the 3DGS-30k baseline, as shown in Figure~\ref{fig:fpsvspsnr}. GF and HAC, despite their efficient compression, exhibit slower rendering speeds due to high encoding/decoding complexity, which improves compression but reduces rendering efficiency. CompGS, however, demonstrates better rendering speeds than 3DGS-30k, HAC, and GF across all benchmark datasets, making it a more balanced approach between compression efficiency and rendering speed.

Our analysis reveals that HAC-lowrate, CompGS (lowrate), and ContextGS (lowrate) achieve the most aggressive compression ratios, while ScaffoldGS, ContextGS (highrate), and HAC-highrate attain the highest fidelity. Although structured compression techniques offer superior storage efficiency compared to unstructured methods, there remains a trade-off between extreme compression and rendering speeds.

%% file: arxiv.bbl
\begin{thebibliography}{100}

\bibitem{kalkofen2008comprehensible}
D.~Kalkofen, E.~Mendez, and D.~Schmalstieg, ``Comprehensible visualization for augmented reality,'' {\em IEEE TVCG}, vol.~15, no.~2, pp.~193--204, 2008.

\bibitem{patney2016towards}
A.~Patney, M.~Salvi, J.~Kim, A.~Kaplanyan, C.~Wyman, N.~Benty, D.~Luebke, and A.~Lefohn, ``Towards foveated rendering for gaze-tracked virtual reality,'' {\em ACM TOG}, vol.~35, no.~6, pp.~1--12, 2016.

\bibitem{albert2017latency}
R.~Albert, A.~Patney, D.~Luebke, and J.~Kim, ``Latency requirements for foveated rendering in virtual reality,'' {\em ACM Transactions on Applied Perception (TAP)}, vol.~14, no.~4, pp.~1--13, 2017.

\bibitem{gortler2023lumigraph}
S.~J. Gortler, R.~Grzeszczuk, R.~Szeliski, and M.~F. Cohen, ``The lumigraph,'' in {\em Seminal Graphics Papers: Pushing the Boundaries, Volume 2}, pp.~453--464, ACM, 2023.

\bibitem{levoy2023light}
M.~Levoy and P.~Hanrahan, ``Light field rendering,'' in {\em Seminal Graphics Papers: Pushing the Boundaries, Volume 2}, pp.~441--452, ACM, 2023.

\bibitem{buehler2023unstructured}
C.~Buehler, M.~Bosse, L.~McMillan, S.~J. Gortler, and M.~F. Cohen, ``Unstructured lumigraph rendering,'' in {\em Proceedings of the 28th Annual Conference on Computer Graphics and Interactive Techniques, {SIGGRAPH} 2001, Los Angeles, California, USA, August 12-17, 2001} (L.~Pocock, ed.), pp.~425--432, {ACM}, 2001.

\bibitem{snavely2006photo}
N.~Snavely, S.~M. Seitz, and R.~Szeliski, ``Photo tourism: exploring photo collections in 3d,'' in {\em ACM TOG}, pp.~835--846, ACM, 2006.

\bibitem{goesele2007multi}
M.~Goesele, N.~Snavely, B.~Curless, H.~Hoppe, and S.~M. Seitz, ``Multi-view stereo for community photo collections,'' in {\em ICCV}, pp.~1--8, 2007.

\bibitem{mildenhall2020nerf}
B.~Mildenhall, P.~P. Srinivasan, M.~Tancik, J.~T. Barron, R.~Ramamoorthi, and R.~Ng, ``Nerf: Representing scenes as neural radiance fields for view synthesis,'' in {\em ECCV}, pp.~405--421, 2020.

\bibitem{liao2024ov}
G.~Liao, K.~Zhou, Z.~Bao, K.~Liu, and Q.~Li, ``Ov-nerf: Open-vocabulary neural radiance fields with vision and language foundation models for 3d semantic understanding,'' {\em IEEE Transactions on Circuits and Systems for Video Technology}, 2024.

\bibitem{lin2025dynamic}
A.~Lin, Y.~Xiang, J.~Li, and M.~Prasad, ``Dynamic appearance particle neural radiance field,'' {\em IEEE Transactions on Circuits and Systems for Video Technology}, 2025.

\bibitem{sheng2024open}
Z.~Sheng, F.~Liu, M.~Liu, F.~Zheng, and L.~Nie, ``Open-set synthesis for free-viewpoint human body reenactment of novel poses,'' {\em IEEE Transactions on Circuits and Systems for Video Technology}, 2024.

\bibitem{zhu2024dfie3d}
X.~Zhu, J.~Zhou, L.~You, X.~Yang, J.~Chang, J.~J. Zhang, and D.~Zeng, ``Dfie3d: 3d-aware disentangled face inversion and editing via facial-contrastive learning,'' {\em IEEE Transactions on Circuits and Systems for Video Technology}, 2024.

\bibitem{ding2024ray}
J.~Ding, Y.~He, B.~Yuan, Z.~Yuan, P.~Zhou, J.~Yu, and X.~Lou, ``Ray reordering for hardware-accelerated neural volume rendering,'' {\em IEEE Transactions on Circuits and Systems for Video Technology}, 2024.

\bibitem{chen2022tensorf}
A.~Chen, Z.~Xu, A.~Geiger, J.~Yu, and H.~Su, ``Tensorf: Tensorial radiance fields,'' in {\em European Conference on Computer Vision}, pp.~333--350, 2022.

\bibitem{garbin2021fastnerf}
S.~J. Garbin, M.~Kowalski, M.~Johnson, J.~Shotton, and J.~Valentin, ``Fastnerf: High-fidelity neural rendering at 200fps,'' in {\em ICCV}, pp.~14346--14355, 2021.

\bibitem{takikawa2021neural}
T.~Takikawa, J.~Litalien, K.~Yin, K.~Kreis, C.~Loop, D.~Nowrouzezahrai, A.~Jacobson, M.~McGuire, and S.~Fidler, ``Neural geometric level of detail: Real-time rendering with implicit 3d shapes,'' in {\em CVPR}, pp.~11358--11367, 2021.

\bibitem{ali2024elmgs}
M.~S. Ali, S.-H. Bae, and E.~Tartaglione, ``Elmgs: Enhancing memory and computation scalability through compression for 3d gaussian splatting,'' {\em arXiv preprint arXiv:2410.23213}, 2024.

\bibitem{qian20233dgs}
Z.~Qian, S.~Wang, M.~Mihajlovic, A.~Geiger, and S.~Tang, ``3dgs-avatar: Animatable avatars via deformable 3d gaussian splatting,'' {\em arXiv preprint arXiv:2312.09228}, 2023.

\bibitem{lee2023compact}
J.~C. Lee, D.~Rho, X.~Sun, J.~H. Ko, and E.~Park, ``Compact 3d gaussian representation for radiance field,'' {\em arXiv preprint arXiv:2311.13681}, 2023.

\bibitem{fan2023lightgaussian}
Z.~Fan, K.~Wang, K.~Wen, Z.~Zhu, D.~Xu, and Z.~Wang, ``Lightgaussian: Unbounded 3d gaussian compression with 15x reduction and 200+ fps,'' {\em arXiv preprint arXiv:2311.17245}, 2023.

\bibitem{navaneet2023compact3d}
K.~Navaneet, K.~Pourahmadi~Meibodi, S.~Abbasi~Koohpayegani, and H.~Pirsiavash, ``Compgs: Smaller and faster gaussian splatting with vector quantization,'' in {\em European Conference on Computer Vision}, pp.~330--349, Springer, 2024.

\bibitem{girish2023eagles}
S.~Girish, K.~Gupta, and A.~Shrivastava, ``Eagles: Efficient accelerated 3d gaussians with lightweight encodings,'' {\em arXiv preprint arXiv:2312.04564}, 2023.

\bibitem{papantonakis2024reducing}
P.~Papantonakis, G.~Kopanas, B.~Kerbl, A.~Lanvin, and G.~Drettakis, ``Reducing the memory footprint of 3d gaussian splatting,'' {\em Proceedings of the ACM on Computer Graphics and Interactive Techniques}, vol.~7, no.~1, pp.~1--17, 2024.

\bibitem{salman2024trimming}
M.~Salman~Ali, M.~Qamar, S.-H. Bae, and E.~Tartaglione, ``Trimming the fat: Efficient compression of 3d gaussian splats through pruning,'' {\em arXiv e-prints}, pp.~arXiv--2406, 2024.

\bibitem{lee2024safeguardgs}
Y.~Lee, Z.~Zhang, and D.~Fan, ``Safeguardgs: 3d gaussian primitive pruning while avoiding catastrophic scene destruction,'' {\em arXiv preprint arXiv:2405.17793}, 2024.

\bibitem{liu2024efficientgs}
W.~Liu, T.~Guan, B.~Zhu, L.~Ju, Z.~Song, D.~Li, Y.~Wang, and W.~Yang, ``Efficientgs: Streamlining gaussian splatting for large-scale high-resolution scene representation,'' {\em arXiv preprint arXiv:2404.12777}, 2024.

\bibitem{wang2024end}
H.~Wang, H.~Zhu, T.~He, R.~Feng, J.~Deng, J.~Bian, and Z.~Chen, ``End-to-end rate-distortion optimized 3d gaussian representation,'' {\em arXiv preprint arXiv:2406.01597}, 2024.

\bibitem{hanson2024pup}
A.~Hanson, A.~Tu, V.~Singla, M.~Jayawardhana, M.~Zwicker, and T.~Goldstein, ``Pup 3d-gs: Principled uncertainty pruning for 3d gaussian splatting,'' {\em arXiv preprint arXiv:2406.10219}, 2024.

\bibitem{kim2024color}
S.~Kim, K.~Lee, and Y.~Lee, ``Color-cued efficient densification method for 3d gaussian splatting,'' in {\em Proceedings of the IEEE/CVF Conference on Computer Vision and Pattern Recognition}, pp.~775--783, 2024.

\bibitem{lin2024rtgs}
W.~Lin, Y.~Feng, and Y.~Zhu, ``Rtgs: Enabling real-time gaussian splatting on mobile devices using efficiency-guided pruning and foveated rendering,'' {\em arXiv preprint arXiv:2407.00435}, 2024.

\bibitem{kheradmand20243d}
S.~Kheradmand, D.~Rebain, G.~Sharma, W.~Sun, J.~Tseng, H.~Isack, A.~Kar, A.~Tagliasacchi, and K.~M. Yi, ``3d gaussian splatting as markov chain monte carlo,'' {\em arXiv preprint arXiv:2404.09591}, 2024.

\bibitem{zhang2024lp}
Z.~Zhang, T.~Song, Y.~Lee, L.~Yang, C.~Peng, R.~Chellappa, and D.~Fan, ``Lp-3dgs: Learning to prune 3d gaussian splatting,'' {\em arXiv preprint arXiv:2405.18784}, 2024.

\bibitem{fang2024mini}
G.~Fang and B.~Wang, ``Mini-splatting: Representing scenes with a constrained number of gaussians,'' {\em European Conference on Computer Vision}, 2024.

\bibitem{di2025gode}
F.~Di~Sario, R.~Renzulli, M.~Grangetto, A.~Sugimoto, and E.~Tartaglione, ``Gode: Gaussians on demand for progressive level of detail and scalable compression,'' {\em arXiv preprint arXiv:2501.13558}, 2025.

\bibitem{niedermayr2024compressed}
S.~Niedermayr, J.~Stumpfegger, and R.~Westermann, ``Compressed 3d gaussian splatting for accelerated novel view synthesis,'' {\em arXiv preprint arXiv:2401.02436}, 2024.

\bibitem{morgenstern2023compact}
W.~Morgenstern, F.~Barthel, A.~Hilsmann, and P.~Eisert, ``Compact 3d scene representation via self-organizing gaussian grids,'' {\em arXiv preprint arXiv:2312.13299}, 2023.

\bibitem{yang2024spectrally}
R.~Yang, Z.~Zhu, Z.~Jiang, B.~Ye, X.~Chen, Y.~Zhang, Y.~Chen, J.~Zhao, and H.~Zhao, ``Spectrally pruned gaussian fields with neural compensation,'' {\em arXiv preprint arXiv:2405.00676}, 2024.

\bibitem{zhang2024gaussian}
F.~Zhang, T.~Zhang, L.~Zhang, H.~Huang, and Y.~Luo, ``Gaussian-forest: Hierarchical-hybrid 3d gaussian splatting for compressed scene modeling,'' {\em arXiv preprint arXiv:2406.08759}, 2024.

\bibitem{liu2024compgs}
X.~Liu, X.~Wu, P.~Zhang, S.~Wang, Z.~Li, and S.~Kwong, ``Compgs: Efficient 3d scene representation via compressed gaussian splatting,'' {\em arXiv preprint arXiv:2404.09458}, 2024.

\bibitem{lu2023scaffold}
T.~Lu, M.~Yu, L.~Xu, Y.~Xiangli, L.~Wang, D.~Lin, and B.~Dai, ``Scaffold-gs: Structured 3d gaussians for view-adaptive rendering,'' {\em arXiv preprint arXiv:2312.00109}, 2023.

\bibitem{chen2024hac}
Y.~Chen, Q.~Wu, J.~Cai, M.~Harandi, and W.~Lin, ``Hac: Hash-grid assisted context for 3d gaussian splatting compression,'' {\em arXiv preprint arXiv:2403.14530}, 2024.

\bibitem{wang2024contextgs}
Y.~Wang, Z.~Li, L.~Guo, W.~Yang, A.~C. Kot, and B.~Wen, ``Contextgs: Compact 3d gaussian splatting with anchor level context model,'' {\em arXiv preprint arXiv:2405.20721}, 2024.

\bibitem{sun2024f}
X.~Sun, J.~C. Lee, D.~Rho, J.~H. Ko, U.~Ali, and E.~Park, ``F-3dgs: Factorized coordinates and representations for 3d gaussian splatting,'' {\em arXiv preprint arXiv:2405.17083}, 2024.

\bibitem{niemeyer2024radsplat}
M.~Niemeyer, F.~Manhardt, M.-J. Rakotosaona, M.~Oechsle, D.~Duckworth, R.~Gosula, K.~Tateno, J.~Bates, D.~Kaeser, and F.~Tombari, ``Radsplat: Radiance field-informed gaussian splatting for robust real-time rendering with 900+ fps,'' {\em arXiv preprint arXiv:2403.13806}, 2024.

\bibitem{henzler2019escaping}
P.~Henzler, N.~J. Mitra, and T.~Ritschel, ``Escaping plato's cave: 3d shape from adversarial rendering,'' in {\em ICCV}, pp.~9984--9993, 2019.

\bibitem{sitzmann2019deepvoxels}
V.~Sitzmann, J.~Thies, F.~Heide, M.~Nie{\ss}ner, G.~Wetzstein, and M.~Zollhofer, ``Deepvoxels: Learning persistent 3d feature embeddings,'' in {\em CVPR}, pp.~2437--2446, 2019.

\bibitem{pfister2000surfels}
H.~Pfister, M.~Zwicker, J.~Van~Baar, and M.~Gross, ``Surfels: Surface elements as rendering primitives,'' in {\em Proceedings of the 27th annual conference on Computer graphics and interactive techniques}, pp.~335--342, 2000.

\bibitem{zwicker2001surface}
M.~Zwicker, H.~Pfister, J.~Van~Baar, and M.~Gross, ``Surface splatting,'' in {\em Proceedings of the 28th annual conference on Computer graphics and interactive techniques}, pp.~371--378, 2001.

\bibitem{ren2002object}
L.~Ren, H.~Pfister, and M.~Zwicker, ``Object space ewa surface splatting: A hardware accelerated approach to high quality point rendering,'' in {\em Computer Graphics Forum}, no.~3, pp.~461--470, 2002.

\bibitem{botsch2005high}
M.~Botsch, A.~Hornung, M.~Zwicker, and L.~Kobbelt, ``High-quality surface splatting on today's gpus,'' in {\em Proceedings Eurographics/IEEE VGTC Symposium Point-Based Graphics, 2005.}, pp.~17--141, 2005.

\bibitem{10757420}
Z.~Guo, W.~Zhou, L.~Li, M.~Wang, and H.~Li, ``Motion-aware 3d gaussian splatting for efficient dynamic scene reconstruction,'' {\em IEEE Transactions on Circuits and Systems for Video Technology}, pp.~1--1, 2024.

\bibitem{jiang2024vr}
Y.~Jiang, C.~Yu, T.~Xie, X.~Li, Y.~Feng, H.~Wang, M.~Li, H.~Lau, F.~Gao, Y.~Yang, {\em et~al.}, ``Vr-gs: A physical dynamics-aware interactive gaussian splatting system in virtual reality,'' {\em arXiv preprint arXiv:2401.16663}, 2024.

\bibitem{liu2024georgs}
Z.~Liu, J.~Su, G.~Cai, Y.~Chen, B.~Zeng, and Z.~Wang, ``Georgs: Geometric regularization for real-time novel view synthesis from sparse inputs,'' {\em IEEE Transactions on Circuits and Systems for Video Technology}, 2024.

\bibitem{10879794}
J.~Zhang, X.~Li, H.~Zhong, Q.~Zhang, Y.~Cao, Y.~Shan, and J.~Liao, ``Humanref-gs: Image-to-3d human generation with reference-guided diffusion and 3d gaussian splatting,'' {\em IEEE Transactions on Circuits and Systems for Video Technology}, pp.~1--1, 2025.

\bibitem{chabra2020deep}
R.~Chabra, J.~E. Lenssen, E.~Ilg, T.~Schmidt, J.~Straub, S.~Lovegrove, and R.~Newcombe, ``Deep local shapes: Learning local sdf priors for detailed 3d reconstruction,'' in {\em ECCV}, pp.~608--625, 2020.

\bibitem{wang2021learning}
Z.~Wang, J.~Philion, S.~Fidler, and J.~Kautz, ``Learning indoor inverse rendering with 3d spatially-varying lighting,'' in {\em ICCV}, pp.~12538--12547, 2021.

\bibitem{10900457}
J.~Liu, L.~Kong, J.~Yan, and G.~Chen, ``Mesh-aligned 3d gaussian splatting for multi-resolution anti-aliasing rendering,'' {\em IEEE Transactions on Circuits and Systems for Video Technology}, pp.~1--1, 2025.

\bibitem{yan2024gs}
C.~Yan, D.~Qu, D.~Xu, B.~Zhao, Z.~Wang, D.~Wang, and X.~Li, ``Gs-slam: Dense visual slam with 3d gaussian splatting,'' in {\em Proceedings of the IEEE/CVF Conference on Computer Vision and Pattern Recognition}, pp.~19595--19604, 2024.

\bibitem{keetha2024splatam}
N.~Keetha, J.~Karhade, K.~M. Jatavallabhula, G.~Yang, S.~Scherer, D.~Ramanan, and J.~Luiten, ``Splatam: Splat track \& map 3d gaussians for dense rgb-d slam,'' in {\em Proceedings of the IEEE/CVF Conference on Computer Vision and Pattern Recognition}, pp.~21357--21366, 2024.

\bibitem{matsuki2024gaussian}
H.~Matsuki, R.~Murai, P.~H. Kelly, and A.~J. Davison, ``Gaussian splatting slam,'' in {\em Proceedings of the IEEE/CVF Conference on Computer Vision and Pattern Recognition}, pp.~18039--18048, 2024.

\bibitem{yugay2023gaussian}
V.~Yugay, Y.~Li, T.~Gevers, and M.~R. Oswald, ``Gaussian-slam: Photo-realistic dense slam with gaussian splatting,'' {\em arXiv preprint arXiv:2312.10070}, 2023.

\bibitem{huang2024photo}
H.~Huang, L.~Li, H.~Cheng, and S.-K. Yeung, ``Photo-slam: Real-time simultaneous localization and photorealistic mapping for monocular stereo and rgb-d cameras,'' in {\em Proceedings of the IEEE/CVF Conference on Computer Vision and Pattern Recognition}, pp.~21584--21593, 2024.

\bibitem{li2024animatable}
Z.~Li, Z.~Zheng, L.~Wang, and Y.~Liu, ``Animatable gaussians: Learning pose-dependent gaussian maps for high-fidelity human avatar modeling,'' in {\em Proceedings of the IEEE/CVF conference on computer vision and pattern recognition}, pp.~19711--19722, 2024.

\bibitem{hu2024gauhuman}
S.~Hu, T.~Hu, and Z.~Liu, ``Gauhuman: Articulated gaussian splatting from monocular human videos,'' in {\em Proceedings of the IEEE/CVF conference on computer vision and pattern recognition}, pp.~20418--20431, 2024.

\bibitem{lei2024gart}
J.~Lei, Y.~Wang, G.~Pavlakos, L.~Liu, and K.~Daniilidis, ``Gart: Gaussian articulated template models,'' in {\em Proceedings of the IEEE/CVF conference on computer vision and pattern recognition}, pp.~19876--19887, 2024.

\bibitem{yuan2024gavatar}
Y.~Yuan, X.~Li, Y.~Huang, S.~De~Mello, K.~Nagano, J.~Kautz, and U.~Iqbal, ``Gavatar: Animatable 3d gaussian avatars with implicit mesh learning,'' in {\em Proceedings of the IEEE/CVF Conference on Computer Vision and Pattern Recognition}, pp.~896--905, 2024.

\bibitem{hu2025tgavatar}
R.~Hu, X.~Wang, Y.~Yan, and C.~Zhao, ``Tgavatar: Reconstructing 3d gaussian avatars with transformer-based tri-plane,'' {\em IEEE Transactions on Circuits and Systems for Video Technology}, 2025.

\bibitem{huang2024endo}
Y.~Huang, B.~Cui, L.~Bai, Z.~Guo, M.~Xu, and H.~Ren, ``Endo-4dgs: Distilling depth ranking for endoscopic monocular scene reconstruction with 4d gaussian splatting,'' {\em arXiv preprint arXiv:2401.16416}, 2024.

\bibitem{liu2024endogaussian}
Y.~Liu, C.~Li, C.~Yang, and Y.~Yuan, ``Endogaussian: Gaussian splatting for deformable surgical scene reconstruction,'' {\em arXiv preprint arXiv:2401.12561}, 2024.

\bibitem{zhao2024hfgs}
H.~Zhao, X.~Zhao, L.~Zhu, W.~Zheng, and Y.~Xu, ``Hfgs: 4d gaussian splatting with emphasis on spatial and temporal high-frequency components for endoscopic scene reconstruction,'' {\em arXiv preprint arXiv:2405.17872}, 2024.

\bibitem{wang2024endogslam}
K.~Wang, C.~Yang, Y.~Wang, S.~Li, Y.~Wang, Q.~Dou, X.~Yang, and W.~Shen, ``Endogslam: Real-time dense reconstruction and tracking in endoscopic surgeries using gaussian splatting,'' in {\em International Conference on Medical Image Computing and Computer-Assisted Intervention}, pp.~219--229, Springer, 2024.

\bibitem{xie2024physgaussian}
T.~Xie, Z.~Zong, Y.~Qiu, X.~Li, Y.~Feng, Y.~Yang, and C.~Jiang, ``Physgaussian: Physics-integrated 3d gaussians for generative dynamics,'' in {\em Proceedings of the IEEE/CVF Conference on Computer Vision and Pattern Recognition}, pp.~4389--4398, 2024.

\bibitem{liu2024physics3d}
F.~Liu, H.~Wang, S.~Yao, S.~Zhang, J.~Zhou, and Y.~Duan, ``Physics3d: Learning physical properties of 3d gaussians via video diffusion,'' {\em arXiv preprint arXiv:2406.04338}, 2024.

\bibitem{borycki2024gasp}
P.~Borycki, W.~Smolak, J.~Waczy{\'n}ska, M.~Mazur, S.~Tadeja, and P.~Spurek, ``Gasp: Gaussian splatting for physic-based simulations,'' {\em arXiv preprint arXiv:2409.05819}, 2024.

\bibitem{huang2024dreamphysics}
T.~Huang, H.~Zhang, Y.~Zeng, Z.~Zhang, H.~Li, W.~Zuo, and R.~W. Lau, ``Dreamphysics: Learning physical properties of dynamic 3d gaussians with video diffusion priors,'' {\em arXiv preprint arXiv:2406.01476}, 2024.

\bibitem{zhang2024physdreamer}
T.~Zhang, H.-X. Yu, R.~Wu, B.~Y. Feng, C.~Zheng, N.~Snavely, J.~Wu, and W.~T. Freeman, ``Physdreamer: Physics-based interaction with 3d objects via video generation,'' in {\em European Conference on Computer Vision}, pp.~388--406, Springer, 2024.

\bibitem{DBLP:journals/corr/abs-2401-03890}
G.~Chen and W.~Wang, ``A survey on 3d gaussian splatting,'' {\em CoRR}, vol.~abs/2401.03890, 2024.

\bibitem{fei20243d}
B.~Fei, J.~Xu, R.~Zhang, Q.~Zhou, W.~Yang, and Y.~He, ``3d gaussian splatting as new era: A survey,'' {\em IEEE Transactions on Visualization and Computer Graphics}, 2024.

\bibitem{bao20243d}
Y.~Bao, T.~Ding, J.~Huo, Y.~Liu, Y.~Li, W.~Li, Y.~Gao, and J.~Luo, ``3d gaussian splatting: Survey, technologies, challenges, and opportunities,'' {\em IEEE Transactions on Circuits and Systems for Video Technology}, 2025.

\bibitem{wu2024recent}
T.~Wu, Y.-J. Yuan, L.-X. Zhang, J.~Yang, Y.-P. Cao, L.-Q. Yan, and L.~Gao, ``Recent advances in 3d gaussian splatting,'' {\em Computational Visual Media}, pp.~1--30, 2024.

\bibitem{bagdasarian20243dgs}
M.~T. Bagdasarian, P.~Knoll, Y.-H. Li, F.~Barthel, A.~Hilsmann, P.~Eisert, and W.~Morgenstern, ``3dgs. zip: A survey on 3d gaussian splatting compression methods,'' {\em arXiv preprint arXiv:2407.09510}, 2024.

\bibitem{kato2020differentiable}
H.~Kato, D.~Beker, M.~Morariu, T.~Ando, T.~Matsuoka, W.~Kehl, and A.~Gaidon, ``Differentiable rendering: A survey,'' {\em arXiv preprint arXiv:2006.12057}, 2020.

\bibitem{ge20193d}
L.~Ge, Z.~Ren, Y.~Li, Z.~Xue, Y.~Wang, J.~Cai, and J.~Yuan, ``3d hand shape and pose estimation from a single rgb image,'' in {\em Proceedings of the IEEE/CVF conference on computer vision and pattern recognition}, pp.~10833--10842, 2019.

\bibitem{baek2019pushing}
S.~Baek, K.~I. Kim, and T.-K. Kim, ``Pushing the envelope for rgb-based dense 3d hand pose estimation via neural rendering,'' in {\em Proceedings of the IEEE/CVF Conference on Computer Vision and Pattern Recognition}, pp.~1067--1076, 2019.

\bibitem{yan2016perspective}
X.~Yan, J.~Yang, E.~Yumer, Y.~Guo, and H.~Lee, ``Perspective transformer nets: Learning single-view 3d object reconstruction without 3d supervision,'' {\em Advances in neural information processing systems}, vol.~29, 2016.

\bibitem{tulsiani2017multi}
S.~Tulsiani, T.~Zhou, A.~A. Efros, and J.~Malik, ``Multi-view supervision for single-view reconstruction via differentiable ray consistency,'' in {\em Proceedings of the IEEE conference on computer vision and pattern recognition}, pp.~2626--2634, 2017.

\bibitem{kato2018neural}
H.~Kato, Y.~Ushiku, and T.~Harada, ``Neural 3d mesh renderer,'' in {\em Proceedings of the IEEE conference on computer vision and pattern recognition}, pp.~3907--3916, 2018.

\bibitem{sitzmann2020implicit}
V.~Sitzmann, J.~Martel, A.~Bergman, D.~Lindell, and G.~Wetzstein, ``Implicit neural representations with periodic activation functions,'' {\em Advances in neural information processing systems}, vol.~33, pp.~7462--7473, 2020.

\bibitem{oleynikova2016signed}
H.~Oleynikova, A.~Millane, Z.~Taylor, E.~Galceran, J.~Nieto, and R.~Siegwart, ``Signed distance fields: A natural representation for both mapping and planning,'' in {\em RSS 2016 workshop: geometry and beyond-representations, physics, and scene understanding for robotics}, University of Michigan, 2016.

\bibitem{drebin1988volume}
R.~A. Drebin, L.~Carpenter, and P.~Hanrahan, ``Volume rendering,'' {\em ACM Siggraph Computer Graphics}, vol.~22, no.~4, pp.~65--74, 1988.

\bibitem{kopanas2021point}
G.~Kopanas, J.~Philip, T.~Leimk{\"u}hler, and G.~Drettakis, ``Point-based neural rendering with per-view optimization,'' in {\em Computer Graphics Forum}, no.~4, pp.~29--43, 2021.

\bibitem{kerbl20233d}
B.~Kerbl, G.~Kopanas, T.~Leimk{\"u}hler, and G.~Drettakis, ``3d gaussian splatting for real-time radiance field rendering,'' {\em ACM TOG}, vol.~42, no.~4, 2023.

\bibitem{zwicker2001ewa}
M.~Zwicker, H.~Pfister, J.~Van~Baar, and M.~Gross, ``Ewa volume splatting,'' in {\em Proceedings Visualization, 2001. VIS'01.}, pp.~29--538, 2001.

\bibitem{knapitsch2017tanks}
A.~Knapitsch, J.~Park, Q.-Y. Zhou, and V.~Koltun, ``Tanks and temples: Benchmarking large-scale scene reconstruction,'' {\em ACM Transactions on Graphics (ToG)}, vol.~36, no.~4, pp.~1--13, 2017.

\bibitem{hedman2018deep}
P.~Hedman, J.~Philip, T.~Price, J.-M. Frahm, G.~Drettakis, and G.~Brostow, ``Deep blending for free-viewpoint image-based rendering,'' {\em ACM TOG}, vol.~37, no.~6, pp.~1--15, 2018.

\bibitem{barron2022mip}
J.~T. Barron, B.~Mildenhall, D.~Verbin, P.~P. Srinivasan, and P.~Hedman, ``Mip-nerf 360: Unbounded anti-aliased neural radiance fields,'' in {\em CVPR}, pp.~5470--5479, 2022.

\bibitem{bakurov2022structural}
I.~Bakurov, M.~Buzzelli, R.~Schettini, M.~Castelli, and L.~Vanneschi, ``Structural similarity index (ssim) revisited: A data-driven approach,'' {\em Expert Systems with Applications}, vol.~189, p.~116087, 2022.

\bibitem{gong2024gdgs}
Y.~Gong, ``Gdgs: Gradient domain gaussian splatting for sparse representation of radiance fields,'' {\em arXiv preprint arXiv:2405.05446}, 2024.

\bibitem{guenter2012foveated}
B.~Guenter, M.~Finch, S.~Drucker, D.~Tan, and J.~Snyder, ``Foveated 3d graphics,'' {\em ACM transactions on Graphics (tOG)}, vol.~31, no.~6, pp.~1--10, 2012.

\bibitem{wandell1995foundations}
B.~A. Wandell, {\em Foundations of vision.}
\newblock sinauer Associates, 1995.

\bibitem{zhai2024splatloc}
H.~Zhai, X.~Zhang, B.~Zhao, H.~Li, Y.~He, Z.~Cui, H.~Bao, and G.~Zhang, ``Splatloc: 3d gaussian splatting-based visual localization for augmented reality,'' {\em arXiv preprint arXiv:2409.14067}, 2024.

\bibitem{zhou2024drivinggaussian}
X.~Zhou, Z.~Lin, X.~Shan, Y.~Wang, D.~Sun, and M.-H. Yang, ``Drivinggaussian: Composite gaussian splatting for surrounding dynamic autonomous driving scenes,'' in {\em Proceedings of the IEEE/CVF Conference on Computer Vision and Pattern Recognition}, pp.~21634--21643, 2024.

\bibitem{gholami2022survey}
A.~Gholami, S.~Kim, Z.~Dong, Z.~Yao, M.~W. Mahoney, and K.~Keutzer, ``A survey of quantization methods for efficient neural network inference,'' in {\em Low-Power Computer Vision}, pp.~291--326, Chapman and Hall/CRC, 2022.

\bibitem{esser2019learned}
S.~K. Esser, J.~L. McKinstry, D.~Bablani, R.~Appuswamy, and D.~S. Modha, ``Learned step size quantization,'' {\em arXiv preprint arXiv:1902.08153}, 2019.

\bibitem{nascimento2023hyperblock}
M.~G.~d. Nascimento, V.~A. Prisacariu, R.~Fawcett, and M.~Langhammer, ``Hyperblock floating point: Generalised quantization scheme for gradient and inference computation,'' in {\em Proceedings of the IEEE/CVF Winter Conference on Applications of Computer Vision}, pp.~6364--6373, 2023.

\bibitem{ali2024towards}
M.~S. Ali, Y.~Kim, M.~Qamar, S.-C. Lim, D.~Kim, C.~Zhang, S.-H. Bae, and H.~Y. Kim, ``Towards efficient image compression without autoregressive models,'' {\em Advances in Neural Information Processing Systems}, vol.~36, 2024.

\bibitem{witten1987arithmetic}
I.~H. Witten, R.~M. Neal, and J.~G. Cleary, ``Arithmetic coding for data compression,'' {\em Communications of the ACM}, vol.~30, no.~6, pp.~520--540, 1987.

\bibitem{minnen2018joint}
D.~Minnen, J.~Ball{\'e}, and G.~D. Toderici, ``Joint autoregressive and hierarchical priors for learned image compression,'' {\em Advances in neural information processing systems}, vol.~31, 2018.

\bibitem{minnen2020channel}
D.~Minnen and S.~Singh, ``Channel-wise autoregressive entropy models for learned image compression,'' in {\em 2020 IEEE International Conference on Image Processing (ICIP)}, pp.~3339--3343, IEEE, 2020.

\bibitem{li2017efficient}
L.~Li, H.~Li, D.~Liu, Z.~Li, H.~Yang, S.~Lin, H.~Chen, and F.~Wu, ``An efficient four-parameter affine motion model for video coding,'' {\em IEEE Transactions on Circuits and Systems for Video Technology}, vol.~28, no.~8, pp.~1934--1948, 2017.

\bibitem{tancik2022block}
M.~Tancik, V.~Casser, X.~Yan, S.~Pradhan, B.~Mildenhall, P.~P. Srinivasan, J.~T. Barron, and H.~Kretzschmar, ``Block-nerf: Scalable large scene neural view synthesis,'' in {\em Proceedings of the IEEE/CVF Conference on Computer Vision and Pattern Recognition}, pp.~8248--8258, 2022.

\bibitem{tang2022compressible}
J.~Tang, X.~Chen, J.~Wang, and G.~Zeng, ``Compressible-composable nerf via rank-residual decomposition,'' {\em Advances in Neural Information Processing Systems}, vol.~35, pp.~14798--14809, 2022.

\bibitem{zheng2024hpc}
Z.~Zheng, H.~Zhong, Q.~Hu, X.~Zhang, L.~Song, Y.~Zhang, and Y.~Wang, ``Hpc: Hierarchical progressive coding framework for volumetric video,'' in {\em Proceedings of the 32nd ACM International Conference on Multimedia}, pp.~7937--7946, 2024.

\bibitem{di2025boost}
F.~Di~Sario, R.~Renzulli, E.~Tartaglione, and M.~Grangetto, ``Boost your nerf: A model-agnostic mixture of experts framework for high quality and efficient rendering,'' in {\em European Conference on Computer Vision}, pp.~176--192, Springer, 2025.

\bibitem{deng2023compressing}
C.~L. Deng and E.~Tartaglione, ``Compressing explicit voxel grid representations: fast nerfs become also small,'' in {\em Proceedings of the IEEE/CVF Winter Conference on Applications of Computer Vision}, pp.~1236--1245, 2023.

\bibitem{deng2022depth}
K.~Deng, A.~Liu, J.-Y. Zhu, and D.~Ramanan, ``Depth-supervised nerf: Fewer views and faster training for free,'' in {\em Proceedings of the IEEE/CVF Conference on Computer Vision and Pattern Recognition}, pp.~12882--12891, 2022.

\bibitem{he2022density}
Y.~He, X.~Ren, D.~Tang, Y.~Zhang, X.~Xue, and Y.~Fu, ``Density-preserving deep point cloud compression,'' in {\em Proceedings of the IEEE/CVF Conference on Computer Vision and Pattern Recognition}, pp.~2333--2342, 2022.

\bibitem{song2023efficient}
R.~Song, C.~Fu, S.~Liu, and G.~Li, ``Efficient hierarchical entropy model for learned point cloud compression,'' in {\em Proceedings of the IEEE/CVF Conference on Computer Vision and Pattern Recognition}, pp.~14368--14377, 2023.

\end{thebibliography}
